\documentclass[aps,prb,twocolumn,reprint,amsmath,amssymb,nofootinbib,superscriptaddress,floatfix,eqsecnum]{revtex4}
\usepackage{graphicx}
\usepackage{amssymb,bbm,amsmath,braket,indentfirst,bm}
\allowdisplaybreaks 
\usepackage[pdftex]{color} 
\usepackage{hyperref}
\usepackage{color}
\usepackage{ulem} 
\usepackage[paperwidth=8.5in,paperheight=11in,centering,hmargin=2cm,vmargin=2.5cm]{geometry} 

\newcommand{\abs}[1]{\ensuremath \left|#1\right|}

\newcommand{\pri}[1]{\ensuremath #1^{\prime}}

\renewcommand\({\ensuremath \left(} \renewcommand\){\ensuremath
  \right)} \renewcommand\[{\ensuremath \left[}
  \renewcommand\]{\ensuremath \right]}
\def\:={\,\raisebox{0.85pt}{.}\hspace{-2.78pt}\raisebox{2.85pt}{.}\!\!=\,}
\def\=:{\,=\!\!\raisebox{0.85pt}{.}\hspace{-2.78pt}\raisebox{2.85pt}{.}\,}

\begin{document}

\title{Accessing topological order in fractionalized liquids with gapped edges}

\author{Thomas~Iadecola} \affiliation{Physics Department, Boston
  University, Boston, Massachusetts 02215, USA}

\author{Titus~Neupert} \affiliation{Princeton Center for Theoretical
  Science, Princeton University, Princeton, New Jersey 08544, USA}

\author{Claudio~Chamon} \affiliation{Physics Department, Boston
  University, Boston, Massachusetts 02215, USA}

\author{Christopher~Mudry} \affiliation{Condensed Matter Theory Group,
  Paul Scherrer Institute, CH-5232 Villigen PSI, Switzerland}

\date{\today}

\begin{abstract}
We consider manifestations of topological order in
time-reversal-symmetric fractional topological liquids (TRS-FTLs),
defined on planar surfaces with holes.  We derive a formula for the
topological ground state degeneracy of such a TRS-FTL, which applies
to cases where the edge modes on each boundary are fully gapped by
appropriate backscattering terms.  The degeneracy is exact in the
limit of infinite system size, and is given by
$q^{N^{\,}_{\mathrm{h}}}$, where $N^{\,}_{\mathrm{h}}$ is the number
of holes and $q$ is an integer that is determined by the topological
field theory.  When the degeneracy is lifted by finite-size effects,
the holes realize a system of $N^{\,}_{\mathrm{h}}$ coupled spin-like
$q$-state degrees of freedom.  In particular, we provide examples
where $\mathbb Z^{\,}_{q}$ quantum clock models are realized on the
low-energy manifold of states.  We also investigate the possibility of
measuring the topological ground state degeneracy with calorimetry,
and briefly revisit the notion of topological order in $s$-wave BCS
superconductors.

\end{abstract}

\maketitle

\section{Introduction}

The robust ground state degeneracy (GSD) that arises in topologically
ordered systems\ \cite{wen89,wen90,wen91} has been an object of intense
study over the past quarter-century.  Interest in such states of
matter has been motivated in large part by the desire to access
quasiparticles with non-Abelian statistics, whose nontrivial braiding
could be used as a platform for quantum computation.\ \cite{nayak}
Nevertheless, to date there has been no definitive experimental proof
that such non-Abelian quasiparticles exist, nor has there been any
direct observation of topological GSD.

There have been several theoretical proposals for the experimental
detection of topological degeneracy.  One set of proposals for the
(putative) non-Abelian $\nu=5/2$ quantum Hall state focuses on
measuring the contribution of the GSD to the electronic portion of the
entropy at low temperatures.  Observable signatures of this
contribution include the thermopower\ \cite{halperin,yang} and the
temperature dependence of the electrochemical potential and orbital
magnetization.\ \cite{cooper} The thermopower has been measured on
several occasions\ \cite{chickering1,chickering2} with no conclusive
signatures.  Abelian fractional quantum Hall (FQH)
states\ \cite{wen_zee} are also topologically ordered, but the bulk GSD
in these systems is only accessible on closed surfaces (e.g.,~the
torus).  This is unnatural for experiments, which are confined to
finite planar systems, although a recent
proposal\ \cite{barkeshli_oreg_qi} suggests a transport measurement in
a bilayer FQH system that avoids this handicap by effectively altering
the topology of the system.

In this paper, we propose that time-reversal-symmetric fractional
topological liquids (FTLs) may constitute a promising alternative
platform for realizing the topological GSD in experimentally
accessible geometries. FTLs with time-reversal symmetry (TRS) have an
effective description in terms of doubled Chern-Simons (CS), or
so-called BF, theories.\ \cite{freedman_nayak} Examples of
time-reversal-symmetric FTLs with topological order include fractional
quantum spin Hall systems,\ \cite{kane_mele1,kane_mele2,bernevig} 
certain spin liquids,\ \cite{thomale}
Kitaev's toric code,\ \cite{tc} and even the $s$-wave BCS
superconductor.\ \cite{wen91,hansson} In the present work we emphasize
FTLs whose edge states in planar geometries can be completely gapped
without breaking TRS, which is possible when certain criteria are
satisfied.\ \cite{levin_stern,neupert} In these cases, the degenerate
ground state manifold is well separated from excited states and the
GSD on punctured planar surfaces is accessible experimentally.

Our program for this paper is as follows. We first derive a formula
for the GSD of a doubled CS theory defined on a plane with
$N^{\,}_{\mathrm{h}}$ holes, in cases where all helical edge modes are
gapped by appropriate backscattering terms.  This topological
degeneracy increases exponentially with the number of holes, and is
exact in the limit where all holes are infinitely large and infinitely
far apart.  We then consider finite-sized systems, where the
degeneracy is split exponentially by quasiparticle tunneling
processes.  In this setting, we argue that the holes themselves
realize an effective spin-like system, whose Hilbert space consists of
what was formerly the degenerate ground state manifold.  We then
examine calorimetry as a possible experimental probe of the
degeneracy.  We argue that, for suitable materials, the contribution
of the GSD to the low-temperature heat capacity could be observed
experimentally, even in the presence of the expected phononic and
electronic backgrounds.  Finally, we also briefly revisit the notion
of topological order in $s$-wave superconductors, which was suggested
by Wen\cite{wen91} and investigated in detail by Hansson et al.~in
Ref.~\onlinecite{hansson}.  We argue that, for a thin-film
superconductor with (3+1)-dimensional electromagnetism, there is
indeed a ground state degeneracy, which is related to flux
quantization.  However, this degeneracy is lifted in a power-law
fashion, rather than exponentially, and is therefore not topological
in the canonical sense of
Refs.~\onlinecite{wen89}--\onlinecite{wen91}.

\section{The topological degeneracy}
\label{sec: The topological degeneracy}

In this section we derive a formula for the ground state degeneracy of
a TRS-FTL with appropriately gapped edges.  We begin with some
preliminary information before moving on to the derivation.

\subsection{Definitions and notation}

A general time-reversal-symmetric doubled Chern-Simons theory in
(2+1)-dimensional space and time has the form\ \cite{neupert}
\begin{subequations}
\label{trscs}
\begin{align}
\mathcal{L}^{\,}_{\mathrm{CS}}&\:= 
\frac{1}{4\pi}\, 
K^{\,}_{\mathsf{ij}}\,
\epsilon^{\mu\nu\rho}\, 
a^{\mathsf{i}}_{\mu}\, 
\partial^{\,}_{\nu}\,
a^{\mathsf{j}}_{\rho} 
+ 
\frac{e}{2\pi}\, 
Q^{\,}_{\mathsf{i}}\,
\epsilon^{\mu\nu\rho}\, 
A^{\,}_{\mu}\, 
\partial^{\,}_{\nu}\,
a^{\mathsf{i}}_{\rho},
\label{dcs}
\end{align}
where $\mathsf{i},\mathsf{j}=1,\ldots,2N$, $\mu,\nu,\rho=0,1,2$, and
summation on repeated indices is implied.  Here, the $2N\times2N$
matrix $K^{\,}_{\mathsf{ij}}$ is symmetric, invertible, and
integer-valued. The fully antisymmetric Levi-Civita tensor
$\epsilon^{\mu\nu\rho}$ appears with the convention
$\epsilon^{012}=1$.  The components $A^{\,}_{\mu}$ of the
electromagnetic gauge potential are restricted to (2+1)-dimensional
space and time, and the vector $\bm{Q}$ has integer entries that
measure the charges of the various CS fields $a^{\mathsf{i}}_{\mu}$ in
units of the electron charge $e$.  The theory contains $N$ Kramers
pairs of CS fields, which transform into one another under the
operation of time-reversal.  We will therefore be particularly
interested in scenarios where the $2N\times 2N$ matrix $K$ has the
following block form, which is consistent with TRS, as was shown in
Ref.~\onlinecite{neupert},
\begin{align}\label{kdef}
K\:=
\begin{pmatrix}
\kappa 
& 
\Delta
\\ 
\Delta^{\mathsf{T}}
&
-\kappa
\end{pmatrix},
\end{align}
where the $N\times N$ matrices $\kappa=\kappa^{\mathsf{T}}$ and
$\Delta =-\Delta^{\mathsf{T}}$.  TRS further imposes that the charge
vector possess the block form (see Ref.~\onlinecite{neupert})
\begin{align}
\bm{Q}\:=
\begin{pmatrix}
\varrho
\\
\varrho
\end{pmatrix}.
\label{qdef}
\end{align}
\end{subequations}
The theory \eqref{trscs} can also be re-expressed in terms of an
equivalent BF theory\ \cite{santos} by defining the linear
transformation
$\tilde{a}^{\mathsf{i}}_{\mu}\:=
R^{\,}_{\mathsf{ij}}\,a^{\mathsf{j}}_{\mu}$,
where
\begin{subequations}\label{bfdef}
\begin{align}\label{rdef}
R\:=
\begin{pmatrix}
\mathbbm{1}
&
\mathbbm{1}
\\ 
\frac{\mathbbm{1}}{2}
& 
-
\frac{\mathbbm{1}}{2}
\end{pmatrix},
\end{align}
with $\mathbbm{1}$ the $N\times N$ identity matrix.  This linear
transformation induces the $K$-matrix and charge vector
\begin{align}
\tilde K &\:= 
(R^{-1})^{\mathsf{T}}\, 
K\, 
R^{-1}=
\begin{pmatrix}
0
&
\varkappa
\\ 
\varkappa^{\mathsf{T}}
&
0
\end{pmatrix},
\\ 
\varkappa&\:=
\kappa
-
\Delta,
\label{eq: def varkappa}
\\ 
\tilde{\bm{Q}}&\:= 
(R^{-1})^{\mathsf{T}}\, 
\bm{Q}
= 
\begin{pmatrix}
\varrho
\\
0 
\end{pmatrix}.
\label{ktrans}
\end{align}
\end{subequations}
Note that the transformation \eqref{rdef} preserves $\det\,K$
[c.f.~Eq.~\eqref{ktrans}].

When defined on a manifold with boundary, the CS theory \eqref{dcs}
has an associated theory of $2N$ chiral bosons $\phi^{\,}_{\mathsf{i}}$
at the edge. In the most generic case, the boundary of the system
consists of a disjoint union of an arbitrary number of edges, each
with a Lagrangian density of the form (in the absence of the gauge
field $A^{\,}_{\mu}$)\ \cite{neupert}
\begin{align}\label{edge}
\mathcal L^{\,}_{\mathrm{E}}&= 
\frac{1}{4\pi}\, 
\left( 
K^{\,}_{\mathsf{ij}}\,
\partial^{\,}_{t}\,\phi^{\,}_{\mathsf{i}}\,
\partial^{\,}_{x}\,\phi^{\,}_{\mathsf{j}} 
- 
V^{\,}_\mathsf{ij}\,
\partial^{\,}_{x}\,\phi^{\,}_{\mathsf{i}}\,
\partial^{\,}_{x}\,\phi^{\,}_{\mathsf{j}} 
\right) 
+ 
\mathcal L^{\,}_{\mathrm{T}},
\end{align}
where $K^{\,}_\mathsf{ij}$ is the same $2N\times2N$ matrix as before
and the positive-definite, real-valued, symmetric
matrix $V^{\,}_\mathsf{ij}$ encodes non-universal information
specific to a particular edge. The Lagrangian density
$\mathcal{L}^{\,}_{\mathrm{T}}$ generically contains all inter-channel
tunneling operators,
\begin{align}
\mathcal{L}^{\,}_{\mathrm{T}}\:= 
\sum_{\bm{T}\in\mathbb{L}} 
U^{\,}_{\bm{T}}(x)\,
\cos
\Big(
\bm{T}^{\mathsf{T}}\,K\,\bm{\phi}(x)
+
\zeta^{\,}_{\bm{T}}(x)
\Big),
\label{tunneling}
\end{align}  
where $\bm{T}$ is a $2N$-dimenisonal integer vector,
$\bm{\phi}^{\mathsf{T}}=(\phi^{\,}_{1}\ \ldots\ \phi^{\,}_{2N})$, and
$\mathbb{L}$ is the set of all tunneling vectors $\bm{T}$ allowed by
TRS and charge conservation (if it holds). The real-valued functions
$U^{\,}_{\bm{T}}(x)$ and $\zeta^{\,}_{\bm{T}}(x)$ encode information
about disorder at the edge and are further constrained to be
consistent with TRS (see Ref.~\onlinecite{neupert}).  When TRS is
imposed, a necessary and sufficient condition for gapping out the
bosonic modes in the edge theory \eqref{edge} is the existence of $N$
$2N$-dimensional vectors $\bm{T}^{\,}_{i}\in\mathbb{L}$ satisfying\ 
\cite{neupert,haldane}
\begin{subequations}
\label{tcriteria}
\begin{align}
& 
\bm{T}^{\mathsf{T}}_{i}\,\bm{Q}= 0,
\quad\forall\ i\indent\text{(charge conservation)},
\label{chargeconservation}
\\
& 
\bm{T}^{\mathsf{T}}_{i}\,K\,\bm{T}^{\,}_{j}= 0,
\quad\forall\ i,j\indent\text{(Haldane criterion)}.
\label{haldanecrit}
\end{align}
\end{subequations}
Strictly speaking, the criterion \eqref{chargeconservation} need not
hold in a general system, such as (for example) in the case of a
superconductor. In this case, one replaces charge conservation with
charge conservation mod 2 (i.e.,~conservation of fermion parity), 
so that $\bm{T}^{\mathsf{T}}_{i}\,\bm{Q}$ is only constrained to be even.  
In the next section, we will focus on cases where the criteria \eqref{tcriteria} 
are satisfied.

\subsection{Gauge invariance in a system with gapped edges}
\label{sec: gluing}

The need for the edge theory \eqref{edge} arises from the failure of
gauge invariance in Chern-Simons theories on manifolds with boundary.
For non-chiral Chern-Simons theories, like those of the form
\eqref{trscs}, the ability to gap out the edge states necessitates an
alternate route to gauge invariance, as we now show.  For simplicity,
we will work on the disk, although analogous results hold for
manifolds with multiple disconnected boundaries.

To proceed, we rewrite the Lagrangian density \eqref{dcs}, in the
absence of the electromagnetic gauge potential $A^{\,}_{\mu}$ (which
we ignore hereafter), in terms of two separate sets of $N$ CS fields 
$\alpha^{\mathtt{i}}$ and $\beta^{\mathtt{i}}$,
\begin{equation}
\begin{split}
\mathcal{L}^{\,}_{\mathrm{CS}}=&\, 
\frac{\epsilon^{\mu\nu\rho}}{4\pi}\,
\left[
\kappa^{\,}_{\mathtt{ij}}\, 
\left(
\alpha^{\mathtt{i}}_{\mu}\,\partial^{\,}_{\nu}\,\alpha^{\mathtt{j}}_{\rho}
-
\beta^{\mathtt{i}}_{\mu}\,\partial^{\,}_{\nu}\,\beta^{\mathtt{j}}_{\rho}
\right) 
\right.  
\\ 
&\, 
\left.  
+ 
\Delta^{\,}_{\mathtt{ij}}
\left(
\alpha^{\mathtt{i}}_{\mu}\,\partial^{\,}_{\nu}\,\beta^{\mathtt{j}}_{\rho}
-
\beta^{\mathtt{i}}_{\mu}\,\partial^{\,}_{\nu}\,\alpha^{\mathtt{j}}_{\rho}
\right)
\right].
\end{split}
\label{annulus}
\end{equation} 
Here, ${\mathtt{i}},\mathtt{j}\in\{1,\ldots,N\}$, and the ``new" CS
fields are defined as $\alpha^{\mathtt{i}}_{\mu}(\bm{x},t)\equiv
a^{\mathtt{i}}_{\mu}(\bm{x},t)$ and
$\beta^{\mathtt{i}}_{\mu}(\bm{x},t)\equiv
a^{{\mathtt{i}}+N}_{\mu}(\bm{x},t)$.  
We define the CS action on the disk $D$ to be
\begin{equation}
S^{\,}_{\mathrm{CS}}\:= 
\int\limits\mathrm{d}t
\int\limits_{D}\mathrm{d}^{2}x\, 
\mathcal{L}^{\,}_{\mathrm{CS}}(\bm{x},t).
\label{scs} 
\end{equation}
Its transformation law under any local gauge transformation of the form
\begin{subequations}
\begin{equation}
\alpha^{\mathtt{i}}_{\mu} 
\mapsto 
\alpha^{\mathtt{i}}_{\mu} 
+
\partial^{\,}_{\mu}\, 
\chi^{\mathtt{i}}_{\alpha}, 
\qquad
\beta^{\mathtt{i}}_{\mu} 
\mapsto 
\beta^{\mathtt{i}}_{\mu} 
+
\partial^{\,}_{\mu}\, 
\chi^{\mathtt{i}}_{\beta},
\end{equation}
where $\mathtt{i}=1,\ldots,N$ and
$\chi^{\mathtt{i}}_{\alpha}$ and $\chi^{\mathtt{i}}_{\beta}$ are
real-valued scalar fields, is
\begin{equation}
S^{\,}_{\mathrm{CS}}\mapsto 
S^{\,}_{\mathrm{CS}}
+
\delta S^{\,}_{\mathrm{CS}}
\end{equation}
with the boundary contribution
\begin{align}
\delta S^{\,}_{\mathrm{CS}}\:=&\, 
\int\limits\mathrm{d}t
\oint\limits_{\partial D}\mathrm{d}x^{\,}_{\mu}\,
\frac{\epsilon^{\mu\nu\rho}}{4\pi}\,
\left[
\kappa^{\,}_{\mathtt{ij}} 
\left( 
\chi^{\mathtt{i}}_{\alpha}\,
\partial^{\,}_{\nu}\, 
\alpha^{\mathtt{j}}_{\rho} 
-
\chi^{\mathtt{i}}_{\beta}\, 
\partial^{\,}_{\nu}\,
\beta^{\mathtt{j}}_{\rho} 
\right) 
\right.\nonumber 
\\ 
&\, 
\left.  
+
\Delta^{\,}_{\mathtt{ij}} 
\left(
\chi^{\mathtt{i}}_{\alpha}\,\partial^{\,}_{\nu}\,\beta^{\mathtt{j}}_{\rho}
-
\chi^{\mathtt{i}}_{\beta}\,\partial^{\,}_{\nu}\,\alpha^{\mathtt{j}}_{\rho}
\right)
\right].
\label{anomaly}
\end{align}
\end{subequations}
Here, the boundary $\partial D$ of the disk $D$ is the circle $S^{1}$, and 
$\mathrm{d}x^{\,}_{\mu}\:=\epsilon^{\,}_{\mu0\sigma}\, \mathrm d\ell^{\sigma}$, 
with $\mathrm{d}\ell^{\sigma}$ the line element along the boundary.

There are two ways to impose gauge invariance in the doubled
Chern-Simons theory $S^{\,}_{\mathrm{CS}}$.  On the one hand, if the criteria
\eqref{tcriteria} do not hold, we must demand that there exist a
gapless edge theory with an action $S^{\,}_{E}$ that transforms as
$S^{\,}_{\mathrm E}\mapsto S^{\,}_{\mathrm E}-\delta S^{\,}_{\mathrm{CS}}$, so that
the total action $S^{\,}_{\mathrm{CS}}+S^{\,}_{\mathrm E}$ is gauge invariant.
On the other hand, if the criteria \eqref{tcriteria} hold, then the
edge fields $\phi^{\,}_\mathsf{i}$ become pinned to the classical
minima of the cosine potentials in $\mathcal{L}^{\,}_{\mathrm{T}}$ for large 
$|U^{\,}_{\bm{T}}(x)|$, and are then no longer dynamical degrees of freedom. In
this case, gauge invariance can be achieved by demanding that the
anomalous term $\delta S^{\,}_{\mathrm{CS}}=0$ identically.  The latter option
can be accomplished by imposing the boundary conditions
\begin{subequations}
\label{gluing} 
\begin{equation}
\chi^{\mathtt{i}}_{\alpha}\vert^{\,}_{\partial D}=
T^{\,}_\mathtt{ij}\,\chi^{\mathtt{j}}_{\beta}\vert^{\,}_{\partial D},
\qquad
\alpha^{\mathtt{i}}_{\mu}\vert^{\,}_{\partial D}=
T^{\,}_\mathtt{ij}\, 
\beta^{\mathtt{j}}_{\mu}\vert^{\,}_{\partial D}, 
\label{gluing1}
\end{equation}
for all $\mathtt{i}=1,\dots,N$, where the invertible $N\times N$
matrix $T$ satisfies the following algebraic criterion:
\begin{equation}
T^{\mathsf{T}}\,
\kappa\,
T 
- 
\kappa 
+ 
T^{\mathsf{T}}\,
\Delta
-
\Delta\,T=0.
\label{gluing2}
\end{equation}
\end{subequations}
One can show that, in order for the boundary conditions
\eqref{gluing1} to be well-defined and consistent with TRS, the matrix
$T$ must have rational entries and satisfy $T^{2}=\mathbbm{1}$ (see
the Appendix).

It is natural to wonder whether different choices of the matrix $T$ in
Eqs.~\eqref{gluing} correspond to different ways of gapping out the
edge theory, i.e.,~to different choices of the set of $N$ linearly
independent tunneling vectors $\bm{T}^{\,}_{i}$ 
($i=1,\dots,N$) that satisfy
Haldane's criterion \eqref{haldanecrit}.  In the Appendix, we argue
that this is indeed the case, although the correspondence need not be
one-to-one.  In particular, while any well-defined choice of the
matrix $T$ implies a particular choice of the set $\{\bm{T}^{\,}_{i}\}$, 
\textit{most} (but not \textit{all}) 
choices of the set $\{\bm{T}^{\,}_{i}\}$ imply a
particular choice of $T$.  In the remainder of this paper, we restrict
our attention to cases where the edge theory is gapped in such a way
that this correspondence holds.

We close this section with the observation that the boundary
conditions \eqref{gluing} can be defined on manifolds with multiple
disconnected boundaries.  For example, the boundary $\partial A$ of
the annulus
\begin{equation}
A\:=[0,\pi]\times S^{1} 
\end{equation}
consists of the disjoint union of two circles ($\partial A = S^{1}\sqcup S^{1}$).  
In this case, one imposes independent boundary conditions of
the form \eqref{gluing1} on each copy of $S^{1}$.  If both edges are
gapped in the same way, then the boundary conditions \eqref{gluing1}
involve the same matrix $T$ on both edges.  It is natural to assume
that this is the case when both boundaries of the annulus separate the
TRS-FTL from vacuum, since both edges have the same symmetries and can
therefore be expected to flow under RG to the same strong-disorder
fixed point with the $N$ most relevant tunneling processes described by
the same set of tunneling vectors $\{\bm T_i\}^{N}_{i=1}$.  We will therefore 
make this assumption in the derivation below.
\subsection{Calculation of the degeneracy}

The ground state degeneracy on the \textit{torus} of a multi-component
Abelian Chern-Simons theory of the form \eqref{dcs} is known on
general grounds to be given by
$|\det\,K|$.\ \cite{wen89,wen_zee,wesolowski} We now present an argument
that, for a doubled CS theory whose $K$-matrix is of
the form \eqref{kdef}, the ground-state degeneracy of the theory on
the \textit{annulus} is given by the formula
\begin{equation}
\text{GSD}=
\sqrt{|\det\,K|}=
\abs{\text{Pf}
\begin{pmatrix}
\Delta
&
\kappa
\\ 
-
\kappa
&
\Delta^{\mathsf{T}}
\end{pmatrix}},
\label{gsd}
\end{equation}
provided that both edges of the annulus are gapped by the same
tunneling terms of the form \eqref{tunneling}, and provided that these
terms are chosen appropriately.  Note that $|\det\,K|$ is the square
of an integer,\ \cite{neupert,santos} so the GSD in these cases is
also an integer.
 
The GSD of non-chiral Chern-Simons theories on manifolds with boundary
depends on the details of how the different edges are gapped (see,
e.g., Ref.~\onlinecite{wang}).  In our argument, this dependence will
manifest itself in different choices of the boundary conditions
\eqref{gluing1} for the bulk Chern-Simons fields, which affect the
counting of the degeneracy.

\begin{figure}[t]
\includegraphics[width=.4\textwidth]{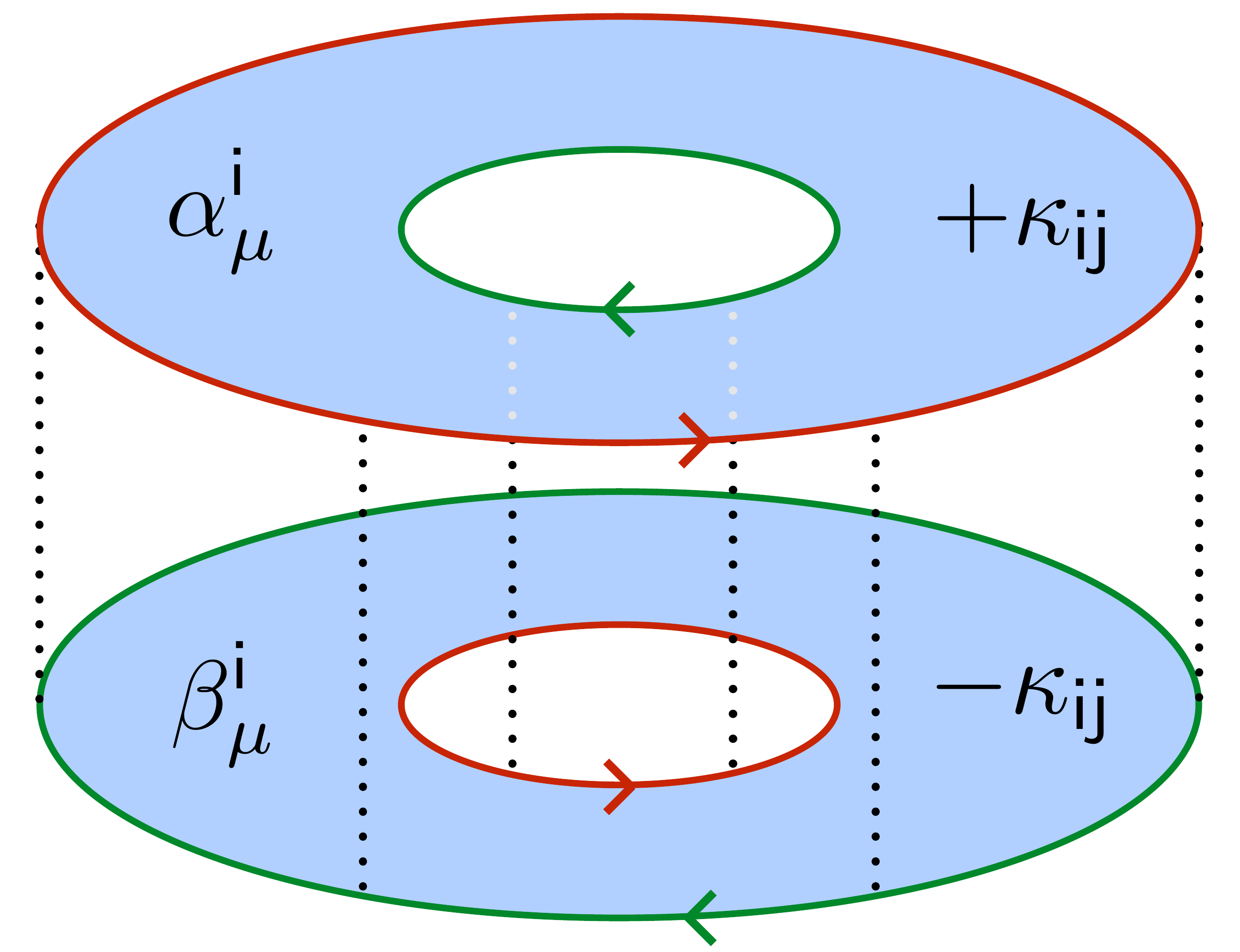}
\caption{(Color online) Gluing argument for the special case
  $\Delta=0$.  In this case, the CS theory consists of two independent
  copies, with equal and opposite $K$-matrices.  The tunneling
  processes (dotted lines) that gap out each pair of
  counterpropagating edge modes couple the two annuli, and the
  conditions \eqref{gluing} ensure that the two copies of the theory
  can be consistently ``glued" together.
\label{fig:gluing}
        }
\end{figure}

Using these boundary conditions, it is possible to show that
Eq.~\eqref{gsd} follows in much the same way as its counterpart on the
torus, so long as both edges of the annulus are gapped by the same
tunneling terms of the form \eqref{tunneling}.  Before proceeding with
the full argument, we first provide an intuitive picture of why this
is, for the case where $\Delta=0$ in Eq.~\eqref{kdef}.  In this case,
Eq.~\eqref{annulus} describes two decoupled CS liquids, one with
$K$-matrix $\kappa$ and the other with $K$-matrix $-\kappa$.  We can
imagine that the two CS liquids live on separate copies of the annulus
$A$, which are coupled by the tunneling processes that gap out the
edges.  The conditions in Eq.~\eqref{gluing} ensure that the two
coupled annuli can be ``glued" together into a single surface, on
which lives a composite CS theory with a GSD given by $|\det \kappa|$
(see Fig.~\ref{fig:gluing}).  Remarkably, these gluing conditions are
also sufficient to treat cases where $\Delta\neq 0$, as we now show.

\subsubsection{Wilson loops, large gauge transformations, and their algebras}

Suppose that we are given a doubled Chern-Simons theory on the annulus
of the form \eqref{trscs}, and that both edges of the annulus are
fully gapped by identical tunneling terms of the form
\eqref{tunneling}.  Let us further impose boundary conditions of the
form \eqref{gluing} at each edge, with the matrix $T$ chosen
appropriately (see the Appendix).  We can now use these boundary
conditions, arising as they do from the need to cancel the anomalous
boundary term \eqref{anomaly}, to construct Wilson loop operators,
which can in turn be used to determine the dimension of the ground
state subspace.

To do this, we first perform a change of basis on the CS Lagrangian
\eqref{dcs} by defining the linear combinations
\begin{subequations}
\begin{align}\label{change basis}
\begin{split}
a^{\mathtt{i}}_{+,\mu}&\:=
T^{\,}_{\mathtt{ij}}\,\alpha^{\mathtt{j}}_{\mu}
+
\beta^{\mathtt{i}}_{\mu}\\
a^{\mathtt{i}}_{-,\mu}&\:=
\frac{1}{2}
\left(
\alpha^{\mathtt{i}}_{\mu}
-
T^{\,}_{\mathtt{ij}}\, 
\beta^{\mathtt{j}}_{\mu}
\right),
\end{split}
\end{align}
where $\mathtt{i}=1,\dots, N$.  
In terms of these fields, the transformed CS Lagrangian reads
\begin{align}
\mathcal{L}^{\,}_{\mathrm{CS}}=& 
\frac{\epsilon^{\mu\nu\rho}}{4\pi}\, 
\(
\varkappa^{\,}_{\mathtt{ij}}\, 
a^{\mathtt{i}}_{+,\mu}\,
\partial^{\,}_{\nu}\, 
a^{\mathtt{j}}_{-,\rho} 
+
\varkappa^{\mathsf{T}}_{\mathtt{ij}}\,
a^{\mathtt{i}}_{-,\mu}\, 
\partial^{\,}_{\nu}\, 
a^{\mathtt{j}}_{+,\rho}
\right.\nonumber\\
&\qquad\left.+
\widetilde\varkappa^{\,}_\mathtt{ij}\, 
a^{\mathtt{i}}_{-,\mu}\, 
\partial^{\,}_{\nu}\, 
a^{\mathtt{j}}_{-,\rho}
\),
\end{align}
where we have defined the $N\times N$ matrices
\begin{align}
\varkappa&\:= 
\kappa\, 
T 
-
T^{\mathsf{T}}\,
\Delta T,
\\ 
\widetilde\varkappa &\:=
\kappa 
-
\Delta\, 
T
-
T^{\mathsf{T}}\,
\Delta^{\mathsf{T}}
-
T^{\mathsf{T}}\,
\kappa T.
\end{align}
Before we continue, note that the linear transformation defined by
Eq.~\eqref{change basis} has determinant $\pm 1$, so that this change
of basis leaves $|\det K|$ invariant.  Consequently, we have that
\begin{align}
|\det K|=
|\det\varkappa|^{2}.
\end{align}  
Furthermore, observe that, in the case $T=\mathbbm{1}$, the matrix
$\varkappa$ above coincides with the one defined in 
Eq.~\eqref{eq: def  varkappa}.  
For reasons that will be made clear below, we restrict
our attention to cases where the matrix $T$ can be chosen such that
the matrix $\varkappa$ has integer entries.
\end{subequations}

\begin{figure}[t]
\includegraphics[width=.33\textwidth]{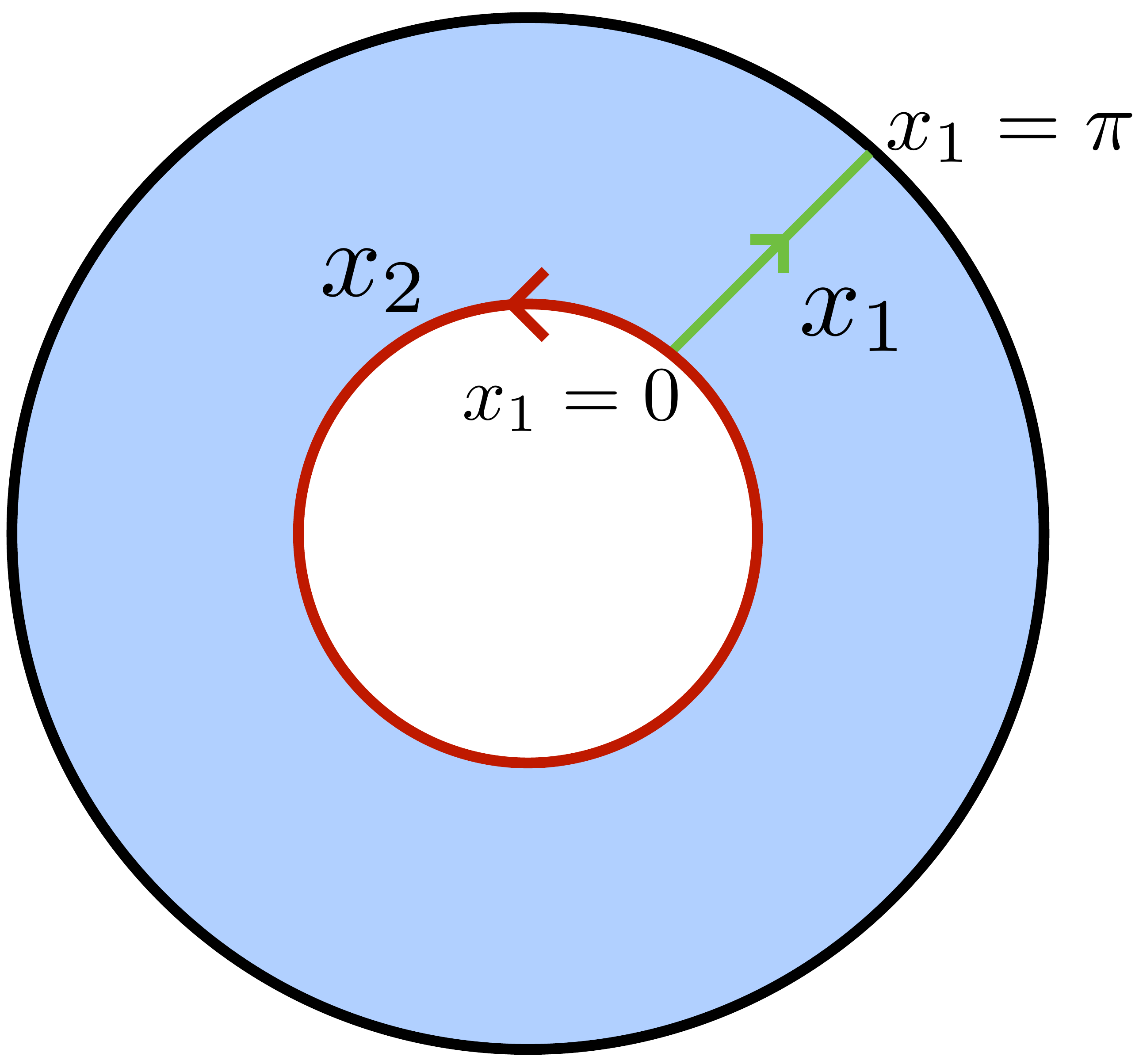}
\caption{(Color online) 
Coordinate system on the annulus
$A=[0,\pi]\times S^{1}$.  
The inner boundary is at $x^{\,}_{1}=0$, while the
outer boundary is at $x^{\,}_{1}=\pi$.  
The coordinate $x^{\,}_{2}$ is defined on the circle $S^{1}$.
\label{fig:annulus}
        }
\end{figure}

In this new basis, the gluing conditions \eqref{gluing} become
Dirichlet boundary conditions on the $(-)$ fields,
\begin{equation}\label{dirichlet}
\chi^{\mathtt{i}}_{-}\vert^{\,}_{\partial A}=0, 
\qquad
a^{\mathtt{i}}_{-,\mu}\vert^{\,}_{\partial A}=0,
\end{equation}
for $\mathtt{i}=1,\ldots,N$.  Rewriting the Lagrangian density in the
gauge $a^{\mathtt{i}}_{\pm,0}=0$ (this can be done using a gauge
transformation obeying the gluing conditions), we obtain
\begin{subequations}
\begin{equation}
\begin{split}
\mathcal{L}^{\,}_{\mathrm{CS}}=&\, 
\frac{1}{4\pi}\,
\left[
\varkappa^{\,}_\mathtt{ij} 
\left( 
a^{\mathtt{i}}_{+,2}\,
\partial^{\,}_{0}\,
a^{\mathtt{j}}_{-,1} 
- 
a^{\mathtt{i}}_{+,1}\, 
\partial^{\,}_{0}\,
a^{\mathtt{j}}_{-,2} 
\right) 
\right.  
\\ 
&\, 
\left.  
+
\varkappa^{\mathsf{T}}_\mathtt{ij}\, 
\left( 
a^{\mathtt{i}}_{-,2}\,
\partial^{\,}_{0}\, 
a^{\mathtt{j}}_{+,1} 
- 
a^{\mathtt{i}}_{-,1}\,
\partial^{\,}_{0}\, 
a^{\mathtt{j}}_{+,2} 
\right)
\right.
\\
&\, 
\left.  
+
\widetilde\varkappa_\mathtt{ij}\, 
\left( 
a^{\mathtt{i}}_{-,2}\,
\partial^{\,}_{0}\, 
a^{\mathtt{j}}_{-,1} 
- 
a^{\mathtt{i}}_{-,1}\,
\partial^{\,}_{0}\, a
^{\mathtt{j}}_{-,2} 
\right)
\right]
\end{split}
\end{equation} 
supplemented by the $2N$ constraints arising from the equations of motion for 
$a^{\mathtt{i}}_{0}$
($\mathtt{i}=1,\ldots,N$),
\begin{equation}
\partial^{\,}_{1}\, 
a^{\mathtt{i}}_{+,2} 
- 
\partial^{\,}_{2}\,
a^{\mathtt{i}}_{+,1}=0, 
\qquad 
\partial^{\,}_{1}\,
a^{\mathtt{i}}_{-,2} 
- 
\partial^{\,}_{2}\, 
a^{\mathtt{i}}_{-,1}=0.
\label{constraints}
\end{equation}
\end{subequations}
The constraints (\ref{constraints}) are met by the decompositions 
\begin{subequations}
\label{decomp}
\begin{align}
& 
a^{\mathtt{i}}_{\pm,1}(x^{\,}_{1},x^{\,}_{2},t)= 
\partial^{\,}_{1}\,
\chi^{\mathtt{i}}_{\pm}(x^{\,}_{1},x^{\,}_{2},t) 
+
\bar{a}^{\mathtt{i}}_{\pm,1}(x^{\,}_{1},t), 
\\ 
&
a^{\mathtt{i}}_{\pm,2}(x^{\,}_{1},x^{\,}_{2},t)= 
\partial^{\,}_{2}\,
\chi^{\mathtt{i}}_{\pm}(x^{\,}_{1},x^{\,}_{2},t) 
+
\bar{a}^{\mathtt{i}}_{\pm,2}(x^{\,}_{2},t),
\end{align}
\end{subequations}
of the CS fields, provided that
$\chi^{\mathtt{i}}_{\pm}(x^{\,}_{1},x^{\,}_{2},t)$
are everywhere smooth functions of $x^{\,}_{1}$ and $x^{\,}_{2}$,
while $\bar{a}^{\mathtt{i}}_{\pm,1}(x^{\,}_{1},t)$ 
and
$\bar{a}^{\mathtt{i}}_{\pm,2}(x^{\,}_{2},t)$ 
are independent of $x^{\,}_{2}$ and $x^{\,}_{1}$,
respectively. Furthermore, the geometry of an annulus is implemented by
the boundary conditions 
\begin{subequations}\label{conditions}
\begin{equation}
\chi^{\mathtt{i}}_{\pm}(x^{\,}_{1},x^{\,}_{2}+2\pi,t)=
\chi^{\mathtt{i}}_{\pm}(x^{\,}_{1},x^{\,}_{2},t)
\end{equation} 
for the fields parametrizing the pure gauge contributions and 
\begin{align}
& 
\chi^{\mathtt{i}}_{-}(0,x^{\,}_{2},t)=
\chi^{\mathtt{i}}_{-}(\pi,x^{\,}_{2},t)=0, 
\\ 
&
\bar{a}^{\mathtt{i}}_{-,1}(0,t)=
\bar{a}^{\mathtt{i}}_{-,1}(\pi,t)=0, 
\\ 
&
\bar{a}^{\mathtt{i}}_{-,2}(x^{\,}_{2},t)\vert^{\,}_{x^{\,}_{1}=0}=
\bar{a}^{\mathtt{i}}_{-,2}(x^{\,}_{2},t)\vert^{\,}_{x^{\,}_{1}=\pi}=
\bar{a}^{\mathtt{i}}_{-,2}(x^{\,}_{2},t)=0,
\end{align}
\end{subequations}
for the gluing conditions.
The coordinate system employed in these definitions is depicted in
Fig.~\ref{fig:annulus}.

The next step is to show that the barred variables decouple from the
remaining (pure gauge) degrees of freedom. This can be done by
inserting the decomposition \eqref{decomp} into the action and using
the boundary conditions \eqref{conditions}.  In the course of this
calculation, the terms containing $\widetilde\varkappa$ that involve
barred variables are found to vanish due to the fact that 
$\bar{a}^{\mathtt{i}}_{-,2}(x^{\,}_{2},t)=0$ for all $x^{\,}_{2}$ and $t$, and to the
periodicity in $x^{\,}_{2}$ of the functions 
$\chi^{\mathtt{i}}_{-}(x^{\,}_{1},x^{\,}_{2},t)$.
We then find an action involving only the matrix $\varkappa$ that
governs the barred variables alone,
\begin{subequations}
\begin{align}
S^{\,}_{\mathrm{top}}= 
\frac{1}{2\pi}\, 
\int\limits\mathrm{d}t\ 
\varkappa^{\,}_{\mathtt{ij}}\, 
A^{\mathtt{i}}_{2}\,
\dot{A}^{\mathtt{j}}_{1},
\label{stop}
\end{align}
where, for all $\mathtt{i}=1,\ldots,N$,
we have defined the global degrees of freedom
\begin{align}
& 
A^{\mathtt{i}}_{1}(t)\:= 
\int\limits\limits_{0}^{\pi}\mathrm{d}x^{\,}_{1}\, 
\bar{a}^{\mathtt{i}}_{-,1}(x^{\,}_{1},t),
\\ 
& 
A^{\mathtt{i}}_{2}(t)\:=
\int\limits\limits_{0}^{2\pi}\mathrm{d}x^{\,}_{2}\,
\bar{a}^{\mathtt{i}}_{+,2}(x^{\,}_{2},t).
\end{align}
\end{subequations}
In Eq.~\eqref{stop},
we employ the notation 
$\dot{A}^{\mathtt{j}}_{1}=
\partial^{\,}_{t}A^{\mathtt{j}}_{1}\equiv
\partial^{\,}_{0}A^{\mathtt{j}}_{1}$.
According to the topological action
(\ref{stop}),
the variable
$\varkappa^{\,}_{\mathsf{ij}}\,A^{\mathtt{i}}_{2}/(2\pi)$
is canonically conjugate to the variable
$A^{\mathtt{j}}_{1}$.
Canonical quantization then gives the equal-time commutation relations
\begin{subequations}
\begin{align}
\[A^{\mathtt{i}}_{1},A^{\mathtt{j}}_{2}\]&=
2\pi\mathrm{i}\, \varkappa^{-1}_{\mathtt{ij}},
\\
\[A^{\mathtt{i}}_{1},A^{\mathtt{j}}_{1}\]&=
\[A^{\mathtt{i}}_{2},A^{\mathtt{j}}_{2}\]=0,
\end{align}
\end{subequations}
for $\mathtt{i},\mathtt{j}=1,\ldots,N$.
We may now define the Wilson loop operators
\begin{subequations}
\begin{equation}
W^{\mathtt{i}}_{1}\:= e^{\mathrm{i}A^{\mathtt{i}}_{1}}, \qquad
W^{\mathtt{i}}_{2}\:= e^{\mathrm{i}A^{\mathtt{i}}_{2}},
\label{wilson}
\end{equation}
whose algebra is found to be
\begin{align}
&
W^{\mathtt{i}}_{1}\,W^{\mathtt{j}}_{2}= 
e^{
-2\pi\mathrm{i}\,
\varkappa^{-1}_{\mathtt{ij}}
  }\, 
W^{\mathtt{j}}_{2}\,
W^{\mathtt{i}}_{1},
\label{eq: algebra between W1 and W2}
\\
&
\[W^{\mathtt{i}}_{1},W^{\mathtt{j}}_{1}\]=
\[W^{\mathtt{i}}_{2},W^{\mathtt{j}}_{2}\]=0.
\label{bfalg1}
\end{align}
\end{subequations}

There is still a set of symmetries that imposes constraints on the
dimension of the Hilbert space associated with $S^{\,}_{\mathrm{top}}$. In
particular, the path integral is invariant under the 
``large gauge transformations"
\begin{equation}
A^{\mathtt{i}}_{1,2}\mapsto 
A^{\mathtt{i}}_{1,2}
+
2\pi
\end{equation}
for any $\mathtt{i}=1,\ldots,N$.
The large gauge transformations are implemented by the operators
\begin{subequations}
\label{bfalg2}
\begin{equation}
U^{\mathtt{i}}_{1}\:= 
e^{
+\mathrm{i}\,
\varkappa^{\,}_{\mathtt{ij}}\,
A^{\mathtt{j}}_{2}
  }, 
\qquad
U^{\mathtt{i}}_{2}\:= 
e^{
-\mathrm{i}\,
\varkappa^{\,}_{\mathtt{ij}}\,
A^{\mathtt{j}}_{1}
  },
\label{bfalg2 a}
\end{equation}
which satisfy the algebra
\begin{align}
&
U^{\mathtt{i}}_{1}\, 
U^{\mathtt{j}}_{2}= 
e^{
-2\pi\mathrm{i}\,
\varkappa^{\,}_{\mathtt{ij}}
  }\, 
U^{\mathtt{j}}_{2}\,
U^{\mathtt{i}}_{1},
\nonumber\\
&
\[
U^{\mathtt{i}}_{1},
U^{\mathtt{j}}_{1}
\]=
\[
U^{\mathtt{i}}_{2},
U^{\mathtt{j}}_{2}
\]=
0,
\label{bfalg2 b}
\end{align}
\end{subequations}
for any $\mathtt{i},\mathtt{j}=1,\ldots,N$.  
Because we require that
$\varkappa$ is an integer matrix, this means that
\begin{equation}
\[
U^{\mathtt{i}}_{1},
U^{\mathtt{j}}_{2}
\]=
\[
U^{\mathtt{i}}_{1},
U^{\mathtt{j}}_{1}
\]=
\[
U^{\mathtt{i}}_{2},
U^{\mathtt{j}}_{2}
\]=
0 
\end{equation}
for all
$\mathtt{i},\mathtt{j}=1,\ldots,N$.  Hence, all 
$U^{\mathtt{i}}_{1}$,
$U^{\mathtt{i}}_{2}$
with $\mathtt{i}=1,\ldots,N$
can be diagonalized simultaneously.  Since any one of $U^{\mathtt{i}}_{1}$
and $U^{\mathtt{i}}_{2}$ generates a transformation
that leaves the path integral invariant, the
vacua of the theory must be eigenstates of any one of
$U^{\mathtt{i}}_{1}$ and $U^{\mathtt{i}}_{2}$ for ${\mathtt{i}}=1,\ldots,N$.

\subsubsection{Dimension of the ground-state subspace}

In order to determine the GSD of the theory, 
it suffices to determine the number of eigenstates of
any one of
$U^{\mathtt{i}}_{1}$ and $U^{\mathtt{i}}_{2}$ for ${\mathtt{i}}=1,\ldots,N$.
To do this, we follow the argument of
Wesolowski et al.,\ \cite{wesolowski} which can be adapted to our case
with only minor modifications. 

First, we define the eigenstates of
any one of
$U^{\mathtt{i}}_{1}$ and $U^{\mathtt{i}}_{2}$ for ${\mathtt{i}}=1,\ldots,N$
by
\begin{align}
U^{\mathtt{i}}_{1}\,
\ket\Psi=
e^{\mathrm{i}\gamma^{\mathtt{i}}_{1}}\,
\ket\Psi,
\qquad
U^{\mathtt{i}}_{2}\,
\ket\Psi=
e^{\mathrm{i}\gamma^{\mathtt{i}}_{2}}\,
\ket\Psi.
\end{align}
Since $A^{\mathtt{i}}_{1}$ and $A^{\mathtt{j}}_{2}$ do not commute, 
we may choose to represent the state $\ket\Psi$
in the basis for which $A^{\mathtt{i}}_{1}$ is diagonal by
\begin{align}
\psi(\{A^{\mathtt{i}}_{1}\})\:= 
\braket{\{A^{\mathtt{i}}_{1}\}|\Psi}.
\label{eq: choice basis 1}
\end{align}
The representation
$\psi(\{A^{\mathtt{i}}_{2}\})$ follows from the representation
$\psi(\{A^{\mathtt{i}}_{1}\})$ by a change of basis to the one
in which $A^{\mathtt{i}}_{2}$ is diagonal.
The large gauge transformations 
(\ref{bfalg2 a})
are represented by
\begin{equation}
U^{\mathtt{i}}_{1}\:= 
e^{2\pi\,\partial/\partial A^{\mathtt{i}}_{1}},
\qquad 
U^{\mathtt{i}}_{2}\:= 
e^{-\mathrm{i}\,\varkappa^{\,}_{\mathtt{ij}}\,A^{\mathtt{j}}_{1}},
\end{equation}
in the basis (\ref{eq: choice basis 1}).
The eigenvalue problem then becomes
\begin{subequations}
\begin{align}
U^{\mathtt{i}}_{1}\, 
\psi(\{A^{\mathtt{i}}_{1}\})\:=&\,
\psi\left(A^{1}_{1},\ldots,A^{\mathtt{i}}_{1}+2\pi,\ldots,A^{N}_{1}\right)
\nonumber\\ 
\equiv&\,
e^{\mathrm{i}\gamma^{\mathtt{i}}_{1}}\,\psi(\{A^{\mathtt{i}}_{1}\}),
\label{eval1}
\\
U^{\mathtt{i}}_{2}\, 
\psi(\{A^{\mathtt{i}}_{1}\})\:=&\,
e^{-\mathrm{i}\,\varkappa^{\,}_{\mathtt{ij}}\,A^{\mathtt{j}}_{1}\,
\psi(\{A^{\mathtt{i}}_{1}\}) }
\nonumber\\
\equiv&\,
e^{\mathrm{i}\gamma^{\mathtt{i}}_{2}}\,
\psi(\{A^{\mathtt{i}}_{1}\}).
\label{eval2}
\end{align}
\end{subequations}
Equation \eqref{eval1} implies that we can write the following series
for $\psi$,
\begin{align}
\psi(\{A^{\mathtt{i}}_{1}\})\equiv 
\psi(\bm{A}^{\,}_{1})=
e^{\mathrm{i}\bm\gamma^{\,}_{1}\cdot\bm{A}^{\,}_{1}/2\pi}\, 
\sum_{\bm n}
d(\bm{n})\, 
e^{\mathrm{i}\bm{n}\cdot\bm{A}^{\,}_{1}},
\label{eq: Fourier expansion psi A's}
\end{align}
where 
$\bm{n}=
(n^{\,}_{1},\ldots,n^{\,}_{N})^{\mathsf{T}}\in\mathbb{Z}^{N}$,
$\bm{A}^{\,}_{1}= 
(A^{1}_{1},\ldots, A^{N}_{1})^{\mathsf{T}}\in\mathbb{R}^{N}$, 
and
$\bm{\gamma}^{\,}_{1}=
(\gamma^{1}_{1},\ldots,\gamma^{N}_{1})^{\mathsf{T}}\in\mathbb{R}^{N}$. 

Second, we seek the constraints on the real-valued coefficients
$d(\bm{n})$ entering the expansion (\ref{eq: Fourier expansion psi A's})
that, as we shall demonstrate, 
fix the dimension of the ground-state subspace.  
To this end, we extract from the $N\times N$
matrix $\varkappa$ that was defined in Eq.~(\ref{eq: def varkappa})
the family
\begin{subequations}
\begin{equation}
\varkappa\=:
\begin{pmatrix}
\bm{k}^{\,\mathsf{T}}_{1} 
\\ 
\vdots 
\\ 
\bm{k}^{\,\mathsf{T}}_{N}
\end{pmatrix}
\end{equation}
of $N$ vectors from $\mathbb{Z}^{N}$ and from its inverse
$\varkappa^{-1}$ the family
\begin{equation}
\varkappa^{-1}\=:
\begin{pmatrix}
\bm{\ell}^{\,}_{1}
&
\ldots
&
\bm{\ell}^{\,}_{N}
\end{pmatrix}
\end{equation}
of $N$ vectors from $\mathbb{Q}^{N}$.  By construction, these vectors
satisfy
\begin{equation}
\bm{k}^{\,}_{\mathtt{i}}\cdot\bm{\ell}^{\,}_{\mathtt{j}}=
\delta^{\,}_{\mathtt{ij}}.
\end{equation}
\end{subequations} 
Using these vectors, we observe that inserting the series
\eqref{eq: Fourier expansion psi A's} 
into the left-hand side of Eq.~\eqref{eval2} gives
\begin{align}
U^{\mathtt{i}}_{2}\, 
\psi(\bm{A}^{\,}_{1})=&\,
e^{\mathrm{i}\bm{\gamma}^{\,}_{1}\cdot\bm{A}^{\,}_{1}/(2\pi)}\,
e^{-\mathrm{i}\bm{k}^{\,}_{\mathtt{i}}\cdot\bm{A}^{\,}_{1}}\,
\sum_{\bm{n}} 
d(\bm{n})\, 
e^{\mathrm{i}\bm{n}\cdot\bm{A}^{\,}_{1}}
\nonumber\\ 
=&\,
e^{\mathrm{i}\bm{\gamma}^{\,}_{1}\cdot\bm{A}^{\,}_{1}/(2\pi)}\,
\sum_{\bm{n}} d(\bm{n}+\bm{k}^{\,}_{\mathtt{i}})\,
e^{\mathrm{i}\bm{n}\cdot\bm{A}^{\,}_{1}} 
\nonumber\\ 
=&\,
e^{\mathrm{i}\gamma_{2}^{\mathtt{i}}}\, \psi(\bm{A}^{\,}_{1}),
\end{align}
which implies
\begin{align}
d(\bm{n}+\bm{k}^{\,}_{\mathtt{i}})=
e^{\mathrm{i}\gamma^{\mathtt{i}}_{2}}\, 
d(\bm{n})
\label{eq: constraint on d of n}
\end{align}
for all ${\mathtt{i}}=1,\ldots,N$. The constraint
(\ref{eq: constraint on d of n})
is automatically satisfied by demanding that
\begin{subequations}
\label{eq: Ansatz for d of n}
\begin{align}
d(\bm{n})=
e^{\mathrm{i}\bm{\gamma}^{\,}_{2}\cdot(\varkappa^{-1})^{\mathsf{T}}\bm{n}}\,
\tilde{d}(\bm{n})
\label{eq: Ansatz for d of n a}
\end{align}
with
\begin{align}
\tilde{d}(\bm{n})= 
\tilde{d}(\bm{n}+\bm{k}^{\,}_{\mathtt{i}}),
\label{eq: Ansatz for d of n b}
\end{align}
since
\begin{align}
\bm{\gamma}^{\,}_{2} 
\cdot 
(\varkappa^{-1})^{\mathsf{T}}\,
\bm{k}^{\,}_{\mathtt{i}}= 
\gamma^{\mathtt{j}}_{2}
(\bm\ell^{\,}_{\mathtt{j}}\cdot\bm{k}^{\,}_{\mathtt{i}})=
\gamma^{\mathtt{i}}_{2}.
\end{align}
\end{subequations}
Hence, insertion of
(\ref{eq: Ansatz for d of n a})
into the expansion 
(\ref{eq: Fourier expansion psi A's}) 
that solves the eigenvalue problem
(\ref{eval1})
gives the expansion
\begin{align}
\psi(\bm{A}^{\,}_{1})=
e^{\mathrm{i}\bm{\gamma}^{\,}_{1}\cdot\bm{A}^{\,}_{1}/(2\pi)}
\sum_{\bm{n}}
e^{\mathrm{i}\bm{\gamma}^{\,}_{2}\cdot(\varkappa^{-1})^{\mathsf{T}}\bm{n}}\,
\tilde{d}(\bm{n})\, e^{\mathrm{i}\bm{n}\cdot\bm{A}^{\,}_{1}}
\end{align}
that solves the eigenvalue problem
(\ref{eval2}).

Third, condition \eqref{eq: Ansatz for d of n b} 
implies that the set of vectors $\{\bm{n}\}$ forms a
lattice with basis vectors $\{\bm{k}^{\,}_{\mathtt{i}}\}$. The number
of inequivalent points in the lattice is therefore given by
\begin{align}
r\:= 
\abs{ 
\det
\begin{pmatrix}
\bm{k}^{\,}_{1}&\ldots&\bm{k}^{\,}_{N}
\end{pmatrix}
    }= 
\abs{\det\,\varkappa^{\mathsf{T}}}= 
|\det\,\varkappa|.
\end{align}
This means that we can decompose any $\bm{n}$ as
\begin{align}
\bm{n}= 
\bm{v}^{\,}_{m} 
+ 
p^{\,}_{\mathtt{i}}\,
\bm{k}^{\,}_{\mathtt{i}},
\end{align} 
where $p^{\,}_{\mathtt{i}}\in\mathbb{Z}$ and we have introduced $r$
linearly independent vectors $\bm{v}^{\,}_{m}$. We can therefore
rewrite
\begin{subequations}
\begin{align}
\psi(\bm{A}^{\,}_{1})= 
\sum_{m=1}^{r} 
\tilde{d}^{\,}_{m}\,
f^{\,}_{m}(\bm{A}^{\,}_{1}),
\end{align}
where
\begin{equation}
\tilde{d}^{\,}_{m}\:=
\tilde{d}(\bm{v}^{\,}_{m}
+
p^{\,}_{\mathtt{i}}\,\bm{k}^{\,}_{\mathtt{i}})=
\tilde{d}(\bm{v}^{\,}_{m}),
\end{equation}
and
\begin{equation}
\begin{split}
f^{\,}_{m}(\bm{A}^{\,}_{1})\:=&\,
e^{\mathrm{i}\bm{\gamma}^{\,}_{1}\cdot\bm{A}^{\,}_{1}/(2\pi)} 
\\ 
&\,
\times 
\!\!\!\!
\sum_{p^{\,}_{1},\ldots,p^{\,}_{N}}
\!\!\!\!
e^{
  \mathrm{i}\bm{\gamma}^{\,}_{2}\cdot(\varkappa^{-1})^{\mathsf{T}} \(
  \bm{v}^{\,}_{m}+p^{\,}_{\mathtt{i}}\,\bm{k}^{\,}_{\mathtt{i}}, \)
}\,
e^{\mathrm{i}(\bm{v}^{\,}_{m}+p^{\,}_{\mathtt{i}}\,\bm{k}^{\,}_{\mathtt{i}})\cdot\bm{A}^{\,}_{1}}.
\end{split}
\end{equation}
\end{subequations}
Since any $\psi(\bm{A}^{\,}_{1})$ in the ground-state manifold can be
written in this way, we have demonstrated that there are
$r=|\det\,\varkappa|$ linearly independent ground-state wavefunctions
$f^{\,}_{m}(\bm{A}^{\,}_{1})$ in the topological Hilbert space. In
other words, we have shown that
\begin{align}
\begin{split}
\text{GSD}&= 
|\det\,\varkappa|=
\sqrt{|\det K|},
\end{split}
\label{proof gsd}
\end{align}
with $K$ defined in Eq.~\eqref{kdef}.  This is precisely the result
advertised in Eq.~\eqref{gsd}.  Note that because $\varkappa$ is an
integer-valued matrix, it has an integer-valued determinant.
Consequently, $\sqrt{|\det K|}=|\det\varkappa|$ is an integer.

\subsubsection{Generalization to manifolds with multiple holes}

It is instructive to consider generalizing these arguments to the case
of a system with the topology of an $N^{\,}_{\mathrm{h}}$-punctured
disk. In this generalization, the boundary can be viewed as the
disjoint union of $N^{\,}_{\mathrm{h}}+1$ copies of $S^{1}$. Since
each of these edges is gapped, anomaly cancellation enforces
independent gluing conditions for each copy of $S^{1}$.  In principle,
a different matrix $T$ could be chosen for each boundary.  This could
happen if, for example, different edges are gapped by different sets
of tunneling vectors $\bm{T}$ that enter Eq.~\eqref{tunneling}.  If
this is the case, then it may not be possible to find a linear
transformation of the form \eqref{change basis} such that $N$ of the
CS fields obey Dirichlet boundary conditions on all edges, as in
Eq.~\eqref{dirichlet}.  The remainder of the argument presented here
for counting the degeneracy then breaks down.  Finding an alternative
argument that applies in these cases is an interesting problem for
future work, but is beyond the scope of this paper.

In the case where all boundaries are gapped in the same way, however,
one obtains a set of Wilson loops like those in Eqs.~\eqref{wilson}
for each hole.  [See, e.g., Eqs.~\eqref{wilson2} in the next section.]
Since these sets of Wilson loops are completely independent, one
obtains a degeneracy of size $|\det\,K|^{N^{\,}_{\mathrm{h}}/2}$.

\section{Applications}

\begin{figure}[t]
(a)\includegraphics[width=.3\textwidth]{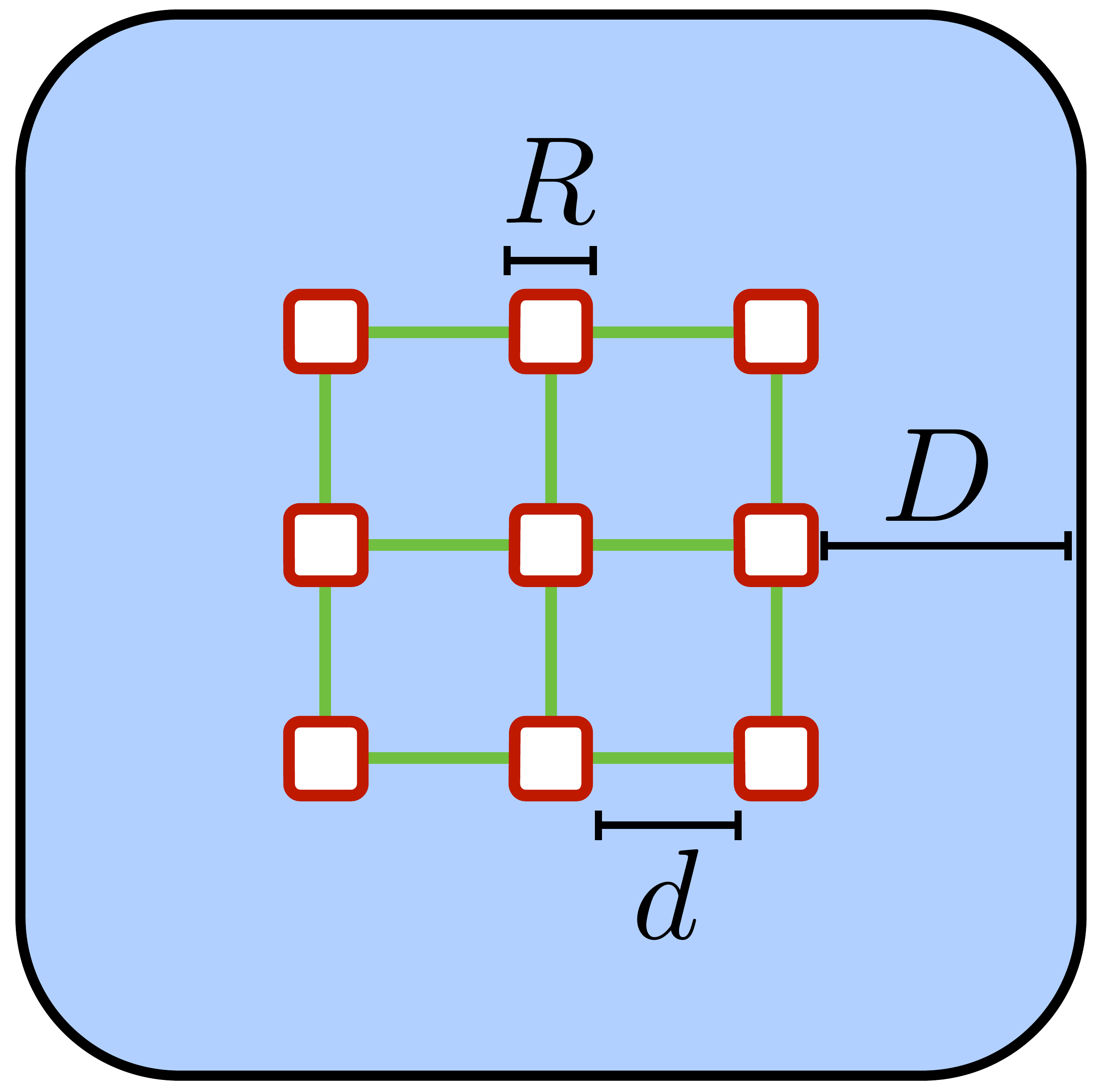}\\
\vspace{1cm} 
(b)\includegraphics[width=.45\textwidth]{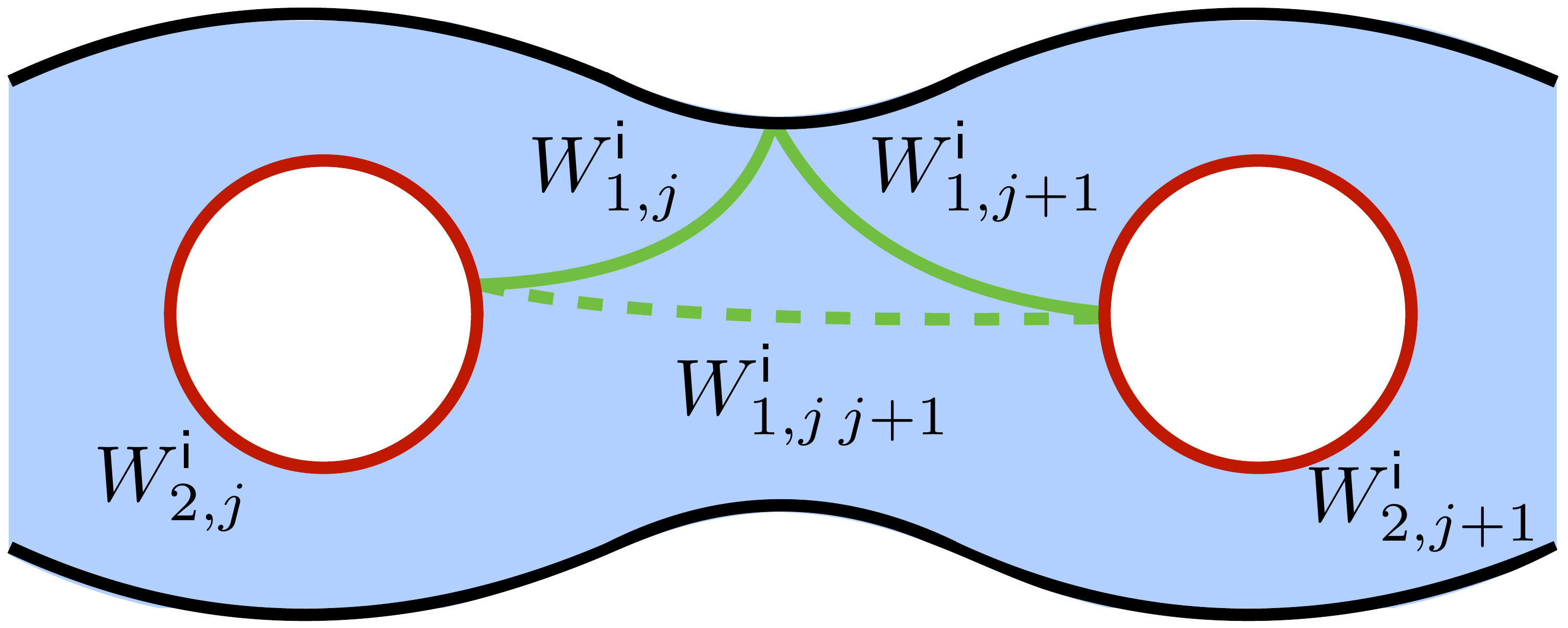}
\caption{(Color online) A punctured TRS-FTL with gapped edges. (a)
Schematic representation of an ``artificial'' spin-like system.  
In the limit $D\gg d,R$, each hole (white square) carries with it a 
$q$-fold topological degeneracy that is split exponentially by
tunneling processes that encircle (red lines) or connect (green
lines) the holes.  (b) Wilson loops defined in Eqs.~\eqref{wilson2}.
The dashed line represents the product of the two Wilson loops above
it, which connects the two holes.
\label{fig: holes}
        }
\end{figure}

With the results of Sec.\ \ref{sec: The topological degeneracy}
in hand, we now explore some
of the consequences of Eq.~\eqref{gsd}.  We begin by examining the
fate of the topological degeneracy in finite-sized systems, before
considering the possibility of using calorimetry to detect
experimental signatures of the degeneracy.  We close the section by
re-evaluating the proposed\ \cite{hansson} topological field theory for
the $s$-wave BCS superconductor in light of the results of this paper.

\subsection{Finite systems: clock models and beyond}\label{parafermions}

On closed manifolds, the topological degeneracy is exact only in the
limit of infinite system size.  This is a result of the fact, pointed
out by Wen and Niu,\ \cite{wen90} that quasiparticle tunneling events
over distances of the order of the system size lift the
topological degeneracy by a splitting
that is exponentially small in the linear size of the system.  This
observation was also confirmed numerically for the case of the
(2+1)-dimensional Abelian Higgs model on the torus by Vestergren et
al.\ in Refs.~\onlinecite{hansson_splitting} and
\onlinecite{vestergren_prb}. A similar splitting occurs for manifolds
with boundary, like those studied in this work.  For a planar system
with many holes, each of which carries a $q$-fold degeneracy (where
$q\:=\sqrt{|\det\, K|}$) in the limit of infinite system size, there
are two kinds of tunneling events that can lift the degeneracy.  These
are (1) tunnelings that encircle a single hole and (2) tunnelings
between boundaries.  Below we argue that, in a finite-sized system
with $N^{\,}_{\mathrm{h}}$ holes, the array of $N^{\,}_{\mathrm{h}}$
coupled $q$-state degrees of freedom can be modeled as a spin-like
system [see Fig.~\ref{fig: holes}(a)].

To see how this arises, we first note that for a system with 
$N^{\,}_{\mathrm{h}}$ 
holes it is possible to define a set of Wilson loops for each hole.  
Analogously to Eqs.~\eqref{wilson}, 
for any $\mathtt{i}=1,\ldots,N$ we define
\begin{subequations}
\label{wilson2}
\begin{align}
&
W^{\mathtt{i}}_{1,j}\:= 
\exp
\Bigg(
\mathrm{i} 
\int\limits_{\mathcal{C}^{\,}_{1,j}}
\mathrm{d}\bm\ell\cdot\bar{\bm{a}}^{\mathtt{i}}_{-}
(\bm{x},t)
\Bigg),
\\
& 
W^{\mathtt{i}}_{2,j}\:= 
\exp
\Bigg(
\mathrm{i}
\oint\limits_{\mathcal{C}^{\,}_{2,j}}
\mathrm{d}\bm\ell\cdot\bar{\bm{a}}^{\mathtt{i}}_{+}
(\bm{x},t)
\Bigg),
\end{align}
where the open curve $\mathcal{C}^{\,}_{1,j}$ connects the $j$-th hole to
the outer boundary, and the closed curve $\mathcal{C}^{\,}_{2,j}$ encircles
the $j$-th hole [see Fig.~\eqref{fig: holes}(b)].  Each set of operators
obeys an independent copy of the algebra \eqref{bfalg1}.  Furthermore,
for any pair of holes $j$ and $k$, the Wilson loop
\begin{equation}
W^{\mathtt{i}}_{1,jk}\:=
W^{\mathtt{i}\,\dag}_{1,j}\,
W^{\mathtt{i}}_{1,k}
\end{equation} 
\end{subequations}
connects these holes. More generally, any number of holes can be
connected by compositions of the Wilson loops defined in Eqs.\
(\ref{wilson2}).
In an infinite system,
the topological protection of the degeneracy 
(\ref{proof gsd})
arises because the Wilson loops defined in Eqs.~\eqref{wilson} 
are nonlocal operators and are
therefore forbidden from entering the Hamiltonian.  In a finite
system, however, the Wilson loops are no longer nonlocal degrees of
freedom and can therefore enter the effective theory. In principle,
all powers and combinations of the Wilson loops are allowed to enter
the effective Hamiltonian
\begin{align}\label{heff}
H^{\,}_{\mathrm{eff}}\:=&\,
\sum_{{\mathtt{i}}=1}^{N}
\sum_{j=1}^{N^{\,}_{\mathrm{h}}}
\Bigg(
h^{\mathtt{i}}_{1,j}\,
W^{\mathtt{i}}_{1,j}
+
h^{\mathtt{i}}_{2,j}\,
W^{\mathtt{i}}_{2,j}
+
\sum_{k=1}^{N^{\,}_{\mathrm{h}}}
J^{\mathtt{i}}_{jk}\,
W^{\mathtt{i}}_{1,jk}
\nonumber\\
&\,
+
\ldots
\Bigg),
\end{align}
where the omitted terms include higher powers of the Wilson loops as
well as all necessary Hermitian conjugates.  In practice, however, all
couplings in $H^{\,}_{\mathrm{eff}}$ are exponentially small in the
shortest available length scale, which limits the tunneling rates.
For example, $J^{\mathtt{i}}_{jk}\propto e^{-c\,d_{jk}/\xi}$, where
$c$ is a constant of order one, $d_{jk}$ is the distance between holes
$j$ and $k$ [see Fig.~\ref{fig: holes}(a)], and $\xi$ is a length scale
associated with quasiparticle tunneling.\ \cite{footnote2}

It is interesting to note that the Hamiltonian $H^{\,}_{\mathrm{eff}}$ 
admits a certain amount of external control -- the holes can be arranged in
arbitrary ways, and the magnitudes of the couplings can be tuned by
changing the length scales $R$, $d^{\,}_{jk}$, and $D$ 
defined in Fig.\ \ref{fig: holes}. 
In particular,
many terms in $H^{\,}_{\mathrm{eff}}$ can be tuned to zero by varying these
length scales.  We will make use of this freedom below.

To illustrate in what sense the effective Hamiltonian \eqref{heff} can be
thought of as a spin-like system, 
we consider a specific class of examples.
In particular, we consider the family of TRS-FTLs defined by
\begin{equation}
K\:=
\begin{pmatrix}q&0\\ 0&-q\end{pmatrix}, 
\qquad
\bm{Q}\:= 
\begin{pmatrix}2\\2 \end{pmatrix},
\label{class}
\end{equation}
where $q$ is an even integer.  One verifies using
Eq.~\eqref{tcriteria} that a single tunneling term of the form
\eqref{tunneling} with $\bm{T} = (1,-1)^{\mathsf{T}}$ is sufficient to
gap out the counterpropagating edge modes without breaking TRS as
defined in Ref.~\onlinecite{neupert}.  (The gluing conditions
\eqref{gluing} can be implemented by the $1\times 1$ gluing ``matrix"
$T=1$.)  In this case, Eq.~\eqref{gsd} predicts a $q$-fold degeneracy
per hole.  To obtain the explicit effective Hamiltonian, we define
\begin{subequations}
\begin{align}
\sigma^{\,}_{j}\:=
W^{\,}_{1,j},
\qquad
\tau^{\,}_{j}\:=
W^{\,}_{2,j},
\end{align}
whose only nonvanishing commutation relations arise from the algebra
[recall Eq.~(\ref{eq: algebra between W1 and W2})]
\begin{align}
\sigma^{\,}_{j}\,
\tau^{\,}_{j}=
e^{-2\pi\mathrm{i}/q}\, 
\tau^{\,}_{j}\,
\sigma^{\,}_{j}.
\end{align}
\end{subequations}
One can check by writing down explicit representations of $\sigma^{\,}_{j}$
and $\tau^{\,}_{j}$ that they also satisfy 
\begin{equation}
\sigma^{q}_{j}=\tau^{q}_{j}=\mathbbm{1}.
\end{equation}
For example, in the case $q=2$ we may use Pauli matrices,
e.g.,
\begin{equation}
\sigma^{\,}_{j}=\sigma^{\,}_{z},
\qquad
\tau^{\,}_{j}=\sigma^{\,}_{x},
\end{equation}
and in the case $q=4$
we may use
\begin{subequations}
\begin{align}
&
\sigma^{\,}_{j}\:=
\text{diag}
\left(
1,
\,e^{-\mathrm{i}\,\pi/2},
\,e^{-\mathrm{i}\,\pi},
\,e^{-\mathrm{i}\,3\pi/2}
\right),
\\ 
&
\tau^{\,}_{j}\:=
\begin{pmatrix}
0&0&0&1
\\ 
1&0&0&0
\\ 
0&1&0&0
\\ 
0&0&1&0
\end{pmatrix}.
\end{align}
\end{subequations}
For a system with $N^{\,}_{\mathrm{h}}$ holes of size $R$ arranged in
a one-dimensional chain with lattice spacing $d$, the effective
Hamiltonian in the limit $D\gg d,R$ (with $D,d,R$ defined in
Fig.\ \ref{fig: holes}) becomes that of a one-dimensional
$\mathbb{Z}^{\,}_{q}$ quantum clock model (see
Ref.~\onlinecite{fendley} and references therein),
\begin{equation}
H^{\,}_{\mathrm{eff}}\:=
\sum_{i=1}^{N^{\,}_{\mathrm{h}}-1}
J^{\,}_{i}\,
\left(
\sigma^{\dag}_{i}\,
\sigma^{\,}_{i+1}
+
\text{H.c.}
\right)
+
\sum_{i=1}^{N^{\,}_{\mathrm{h}}}
h^{\,}_{i}
\left(
\tau^{\,}_{i}
+
\text{H.c.}
\right),
\label{clock}
\end{equation}
where 
$J^{\,}_{i}\propto e^{-c^{\,}_{1}\, d/\xi}$ 
and 
$h^{\,}_{i}\propto e^{-c^{\,}_{2}\, R/\xi}$,
with the real constants $c^{\,}_{1}$ and $c^{\,}_{2}$ of order
unity.  
For simplicity, we have constrained the
couplings $J^{\,}_{i}$ and $h^{\,}_{i}$ to be real, although their magnitude and
sign is allowed to vary from hole to hole (hence the subscripts $i$).
Note that in the above Hamiltonian, terms linear in $\sigma^{\,}_{j}$ do not
appear, as the associated couplings are suppressed by factors of order
$e^{-c^{\,}_{3}\,D/\xi}\ll e^{-c^{\,}_{1}\,d/\xi},\ e^{-c^{\,}_{2}\,R/\xi}$.
Similarly, longer-range two-body terms, as
well as higher powers of the $\sigma^{\,}_{j}$ and $\tau^{\,}_{j}$, are also
omitted, as they correspond to higher-order tunneling processes.

The Hamiltonian of the clock
model \eqref{clock} is invariant under the symmetry operation
\begin{subequations}
\begin{equation}
H^{\,}_{\mathrm{eff}}\mapsto
\mathcal{S}\,
H^{\,}_{\mathrm{eff}}\,
\mathcal{S}^{-1}
\end{equation}
generated by
\begin{equation}
\mathcal{S}\:= 
\prod_{i=1}^{N^{\,}_{\mathrm{h}}}\tau^{\dag}_{i}.
\end{equation}
\end{subequations}
Indeed, under conjugation by 
$\mathcal{S}$,
$\tau^{\dag}_{j}\mapsto\tau^{\dag}_{j}$ and 
$\sigma^{\dag}_{j}\mapsto e^{-2\pi\mathrm{i}/q}\sigma^{\dag}_{j}$ 
for all $j$.  This $\mathbbm{Z}^{\,}_{q}$ 
symmetry can be thought of as a remnant of the
$q^{N^{\,}_{\mathrm{h}}}$-fold topological degeneracy of the TRS-FTL,
which would be present in the limit $d,R,D\to\infty$.

Before closing this section, we point out that quantum clock models
like the one discussed in this section have arisen in various contexts
elsewhere in the recent literature, especially in quantum Hall systems
with defects. \cite{barkeshli_oreg_qi,clarke,vaeziPRX,mongPRX}

\subsection{Probing the topological degeneracy with calorimetry}

In this section, we consider experimental avenues to
detect the topological degeneracy of a punctured TRS-FTL.  We focus
our attention on calorimetry as a possible probe.  In a sample with
$N^{\,}_{\mathrm{h}}$ holes, the ground state degeneracy provides a contribution 
$S^{\,}_{\mathrm{GSD}}=N^{\,}_{\mathrm{h}}\,k^{\,}_{\mathrm{B}}\ln q$, 
where $k^{\,}_{\mathrm{B}}$ is the Boltzmann constant and
$q=\sqrt{\det K}$, to
the total entropy $S^{\,}_{\mathrm{tot}}$.  
If the areal density of holes is kept fixed, then
for a sample of length $L$, we have 
$S^{\,}_{\mathrm{GSD}}\sim L^{2}$ for the topological contribution, 
which is extensive.  This suggests that, were a suitable material to be
discovered, one might be able to detect the topological degeneracy of
a punctured TRS-FTL by measuring its heat capacity.  Such a
measurement is feasible with current technology, as membrane-based 
nanocalorimeters enable the determination of heat capacities $C^{\,}_{V}$ 
in microgram samples (and smaller), to an accuracy of 
$\delta C^{\,}_{V}/C^{\,}_{V}\sim10^{-4}$--$10^{-5}$ 
down to temperatures of order 
$100$ mK.%
\ \cite{garden_review,ong,tagliati1,tagliati2}

We first determine the topological contribution to the heat capacity
for some particular examples.  To do this, we return to the class of
TRS-FTLs defined in Eq.~\eqref{class}.
The heat capacity in this case is easiest to determine from the clock
model of Eq.~\eqref{clock} in the paramagnetic limit $J^{\,}_{i}\to 0$, which
is achieved for $d\gg R$ [see Fig.~\ref{fig: holes}(a)].  Setting $h^{\,}_{i}=h$
for convenience, we see that the clock model can be rewritten, 
after a change of basis, as
\begin{equation}
H^{\,}_{\mathrm{eff}}= 
h\sum_{i=1}^{N^{\,}_{\mathrm{h}}} 
\left(
\sigma^{\,}_{i}
+
\sigma^{\dag}_{i}
\right)=
2h
\sum_{i=1}^{N^{\,}_{\mathrm{h}}} 
\cos\(\frac{2\pi}{q}\,n^{\,}_{i}\),
\end{equation}
where $n^{\,}_{i}=0,\ldots,q-1$.  
Consequently the partition function is given by
\begin{align}
Z= 
\left(
\sum_{n=0}^{q-1}
e^{-2\beta\,h\,\cos(2\pi\,n/q)}
\right)^{N^{\,}_{\mathrm{h}}},
\end{align}
where $\beta\:=1/(k^{\,}_{\mathrm{B}}\,T)$ and $T$ is the temperature.
The topological heat capacity at constant volume, $C^{\mathrm{top}}_{V}$,
is then determined from
the partition function by standard methods.  For example,
\begin{equation} 
C^{\mathrm{top}}_{V}= 
N^{\,}_{\mathrm{h}}\, 
\frac{h^{2}}{k^{\,}_{\mathrm{B}}\,T^{2}}
\times
\begin{cases} 
4\,
\text{sech}^{2}\(\frac{2\,h}{k^{\,}_{\mathrm{B}}\,T}\),
& 
q=2,
\\ 
2\,
\text{sech}^{2}\(\frac{h}{k^{\,}_{\mathrm{B}}\,T}\),
& 
q=4,
\\ 
\frac{
9\,
\cosh\left(\frac{h}{k^{\,}_{\mathrm{B}}\,T}\right)
+
\cosh\left(\frac{3\,h}{k^{\,}_{\mathrm{B}}\,T}\right)
+
8
     }
     {
\left[2\,\cosh\left(\frac{h}{k^{\,}_{\mathrm{B}}\,T}\right)
+
\cosh\left(\frac{2\,h}{k^{\,}_{\mathrm{B}}\,T}\right)\right]^{2}
     },
& 
q=6,
\end{cases}
\end{equation}
and so on.

To date, there has been no experimental realization of a TRS-FTL or
fractional topological insulator.  Since background contributions to
the heat capacity are material-dependent, it is difficult to provide a
precise estimate of the observable effect.  However, we can
nevertheless identify some constraints on the possible materials that
would favor such a measurement.  

To do this, let us estimate the various background contributions 
to the heat capacity of a TRS-FTL. First, we note that, 
because any TRS-FTL must have a gap $\Delta$, the electronic contribution 
$C^{\mathrm{el}}_{V}$ to the heat capacity is
\begin{subequations} 
\begin{align}
C^{\mathrm{el}}_{V}\propto 
\frac{\Delta}{T}\, 
e^{-\eta\,\Delta/(k^{\,}_{\mathrm{B}}\,T)},
\end{align}
where $\eta$ is a constant of order one.  The exponential suppression
of $C^{\mathrm{el}}_{V}$ implies that this contribution is always negligible
at sufficiently small temperatures.  

However, one must also consider
the phononic contribution, which follows a Debye power law at low
temperatures.  This contribution scales with the sample \textit{volume},
which could be three-dimensional if the TRS-FTL is formed in a
heterostructure, as is the case in quantum Hall systems.  This fact,
which was noted in Ref.~\onlinecite{cooper}, poses the greatest
challenge to detecting the topological contribution to the heat
capacity, which scales with the \textit{area} of the two-dimensional
sample.  In principle, however, one may assume that the TRS-FTL lives
in a strictly two-dimensional sample, or at least in a thin film.  In
this case, we have that the phononic contribution $C^{\mathrm{ph}}_{V}$
to the heat capacity is
\begin{equation}
C^{\mathrm{ph}}_{V}\propto  
k^{\,}_{\mathrm{B}}\, (T/T^{\,}_{\mathrm{D}})^{2},
\end{equation}
where $T^{\,}_{\mathrm{D}}$ is the Debye temperature (100 K, say).%
\ \cite{footnote3}
We verified numerically, by simulating a square lattice of masses and
springs, that the presence or absence of holes has little effect on
the phonon spectrum as long as the holes are sufficiently small.  We
therefore expect the Debye law to hold both with and without holes, as
long as one takes into account the excluded volume due to the holes.

\begin{figure}[t]
\hspace{-.35
  cm}
  \includegraphics[width=.5\textwidth]{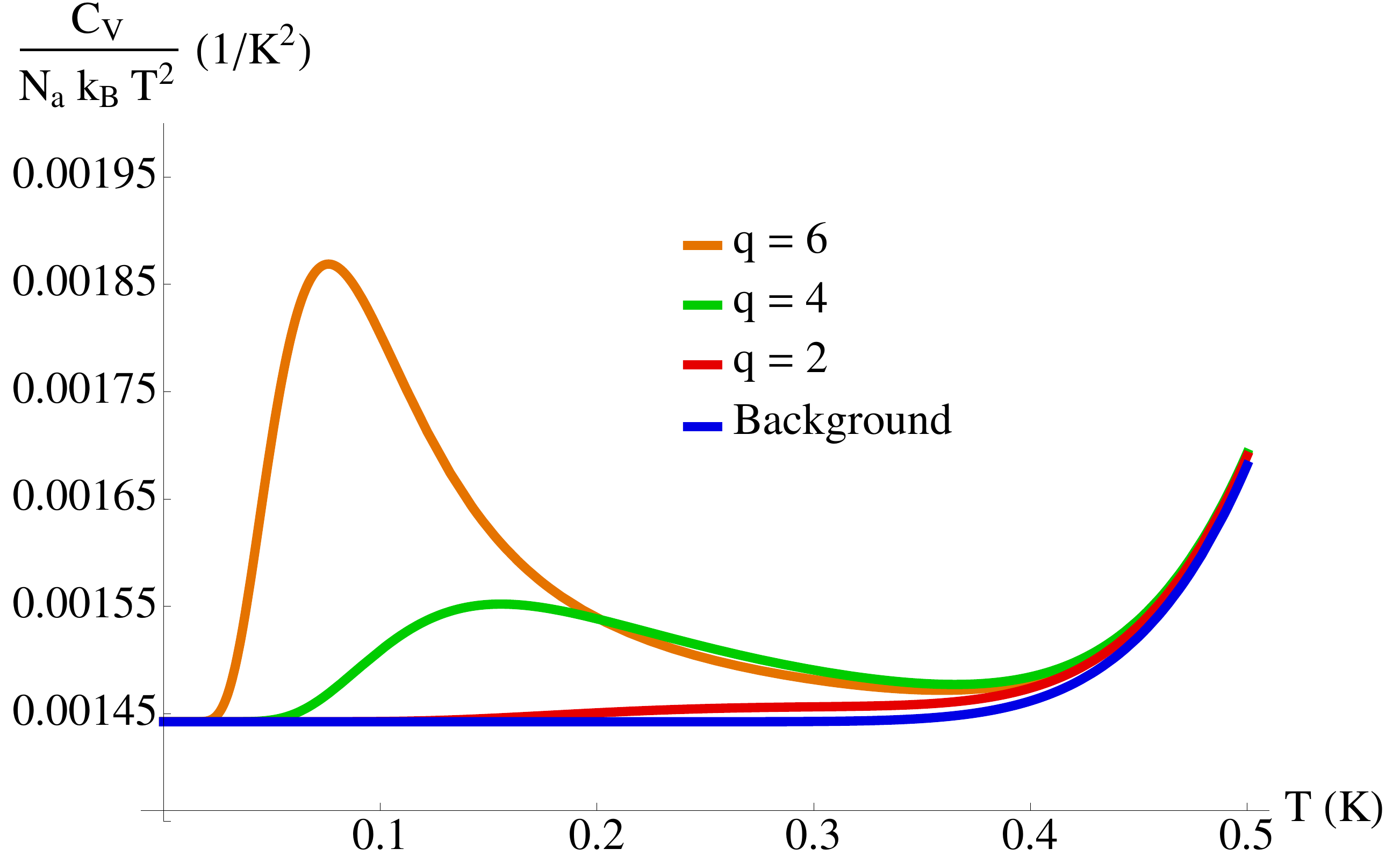}
\caption{(Color online) 
Total heat capacity for a monolayer TRS-FTL
with $N^{\,}_{\mathrm{a}}=10^{14}$.  
The topological contribution is shown
(above background) for $q=2,4,$ and $6$.  
The parameters used for the topological contribution were 
$\nu=5\times 10^{-6}$ ($\sim 22000^{2}$ holes) and 
$h/k^{\,}_{\mathrm{B}}\approx 0.321$ K, which leads to a maximum
excess (for $q=6$) of $\sim 30\%$ over the background (blue curve)
near $T=0.1$ K.
\label{heat_capacity}
        }
\end{figure}

The total heat capacity is obtained by adding the three contributions:
\begin{align}
C^{\,}_{V}(T)= 
N^{\,}_{\mathrm{a}}\,
\left[
C^{\mathrm{top}}_{V}(T)
+
\nu\,
C^{\mathrm{ph}}_{V}(T)
+
\frac{1}{N^{\,}_{\mathrm{a}}}\,
C^{\mathrm{el}}_{V}(T)
\right],
\end{align}
\end{subequations}
where $N^{\,}_{\mathrm a}$ is the number of atoms in the sample and
$\nu\:=N^{\,}_{\mathrm{h}}/N^{\,}_{\mathrm{a}}$ determines the number of holes.  
The above formula
leads to the estimate of the specific heat curve presented in
Fig.~\ref{heat_capacity}.  A square array of $22000$ holes on a side
produces an excess of up to $30\%$ (for $q=6$) on top of the
background at $T=0.1$ K, which is well above the experimental error
$\delta C^{\,}_{V}/C^{\,}_{V}\sim 10^{-4}$.

We now comment on possible difficulties with this measurement.
Perhaps the most important of these is the fact that the energy scales
$J^{\,}_{i}$ and $h^{\,}_{i}$ entering Eq.~\eqref{clock} are unknown.  It may be
possible to circumvent this issue by exploiting the exponential
sensitivity of the couplings to the length scales $R$ and $d$.  For
example, one could prepare samples with $d\gg R$ to eliminate the first
term in Eq.~\eqref{clock}, and compare results for different values of
$R$ to determine whether it is possible to resolve the effect.  As
long as $h\gtrsim 0.1\,\Delta$, it should be possible to tune $R$ such
that the effect is visible.

The presence of disorder in the sample is another potential source of
difficulty, as localized states due to disorder can also contribute to the entropy.
However, intuition from noninteracting systems, where these states
provide a logarithmic correction to the entropy,\cite{footnote_disorder}
suggests that this contribution would be subleading as compared to the
power-law contribution $S_{\rm GSD}\sim L^2$ that we predict for a
fixed areal density of holes.

\subsection{Are superconductors topologically ordered?}

In an insightful paper, it was argued by Hansson et al.\
in Ref.~\onlinecite{hansson}
that ordinary $s$-wave BCS superconductors are topologically ordered.
In fact, it was shown that, when the electromagnetic gauge field is
treated dynamically and confined to (2+1) dimensional space and time, 
the superconductor admits a description in terms of a BF theory like the
one defined in Eqs.~\eqref{bfdef}, with
\begin{align}
\tilde{K}= 
\begin{pmatrix} 
0
&
2
\\ 
2
&
0
\end{pmatrix}.
\label{sc}
\end{align}
Furthermore, it was shown that the edge states that arise when 
the above theory is defined in a finite planar geometry are generically 
gapped by Cooper pair creation terms.  
The proposed theory is consistent with the time-reversal symmetry of the
$s$-wave superconductor and captures the statistical phase of $\pi$
that is acquired by an electron upon encircling a vortex.  This
effective theory, which is the same as that of the $\mathbbm{Z}^{\,}_{2}$
lattice gauge theory in its deconfined phase, predicts a four-fold GSD
on the torus, whose exponential splitting in finite systems was
verified numerically in Refs.~\onlinecite{hansson_splitting} and
\onlinecite{vestergren_prb}.

Since the theory defined by Eq.~\eqref{sc} falls squarely within the
class of theories studied in this paper, it is tempting to draw the
conclusion that the $s$-wave superconductor exhibits a two-fold GSD on
the annulus.  Below we argue that, while this is indeed the case, the
degeneracy is not exponential but power-law in nature, and therefore
is not what one might call a topological degeneracy in the canonical
sense of Refs.~\onlinecite{wen89}--\onlinecite{wen91}.  The reason for this 
is that the topological nature of the superconductor results from the dynamics 
of the electromagnetic gauge field, which, in a real planar superconductor, 
is not confined to the sample itself, but rather extends through all 
three spatial dimensions.  
Consequently, the true electromagnetic gauge field that is present in the 
superconductor can be measured by local external probes.

\begin{figure}[t]
\includegraphics[width=.33\textwidth]{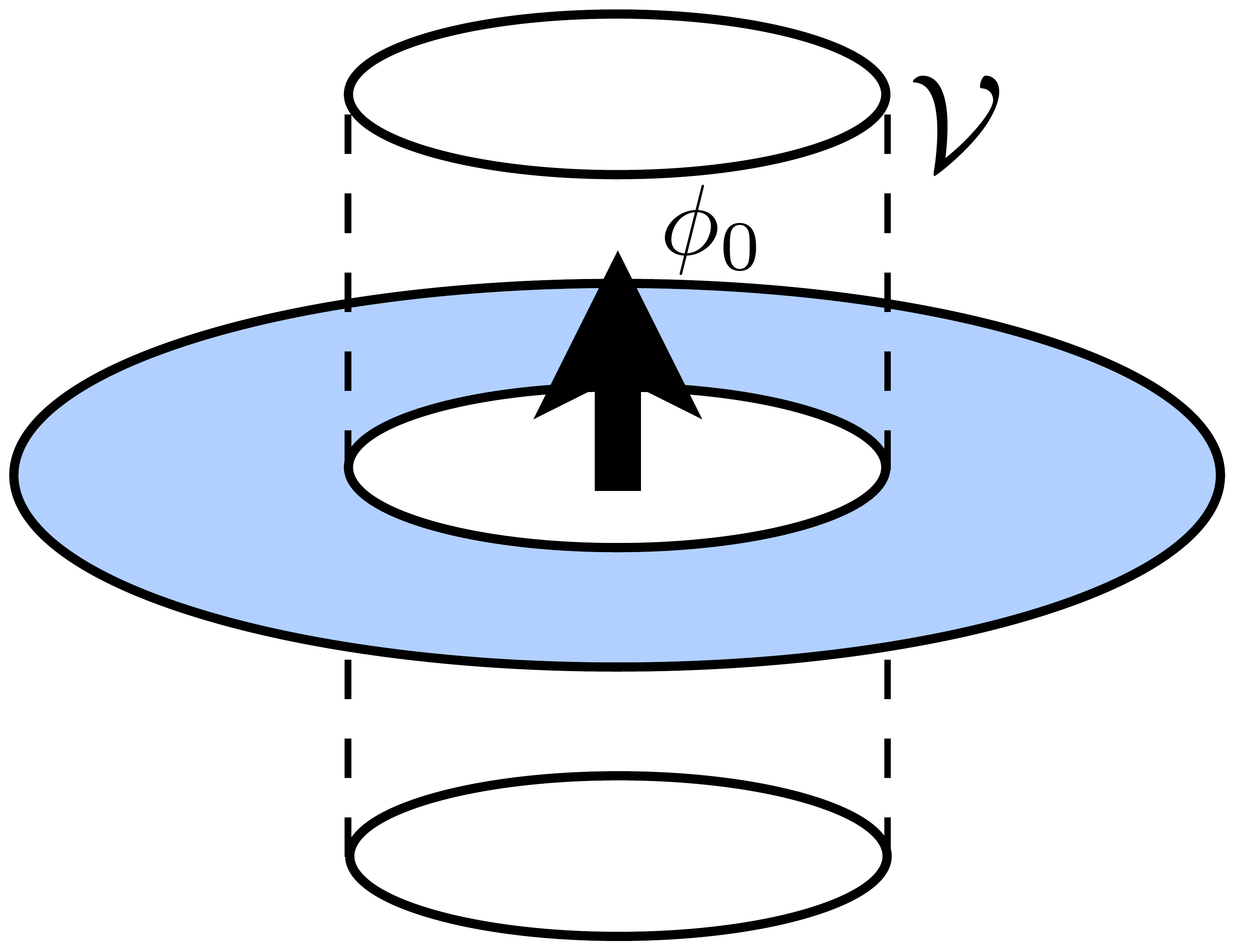}
\caption{(Color online) Trapping a flux quantum inside a
  superconducting ring.  Confining the flux inside the ring costs no
  energy for the electrons inside the superconductor, but there is an
  electromagnetic energy cost obtained by integrating the enclosed
  magnetic field intensity over the interior of the dashed cylinder,
  which we denote $\mathcal V$.
\label{flux}
        }
\end{figure}

To see how this coupling to the environment lifts the degeneracy in a
power-law fashion, let us consider the origin of the two-fold
degeneracy.  Recall that for an annular superconductor (a thin-film
mesoscopic ring, for example), the phase of the superconducting order
parameter winds by $2\pi$ around the hole if a flux quantum
$\phi^{\,}_{0}=h/2e$ is trapped inside.  This indicates that the electronic
spectrum of the superconductor cannot be used to distinguish between
cases where an even ($\phi=0$ mod $\phi^{\,}_{0}$) or odd ($\phi=1$ mod
$\phi^{\,}_{0}$) number of flux quanta penetrate the hole.  This is precisely
the origin of the degeneracy.  However, because the electromagnetic field 
also exists outside the sample, there is an additional
electromagnetic energy cost associated with having a flux quantum
trapped in the hole.  If we assume for simplicity that the flux is
distributed uniformly over the hole (radius $R$) and does not
penetrate into the superconductor, then the energy cost is
proportional to
\begin{equation}
\int\limits_{\mathcal{V}} 
\mathrm{d}^{3}r\, 
|\bm{B}|^{2}=
\frac{\phi^{2}_{0}}{2\pi\,R^{2}}\,
L^{\,}_{z},
\end{equation}
where $\mathcal{V}$ is the interior of the cylinder in
Fig.~\ref{flux}, and $L^{\,}_{z}$ is the height of the cylinder.
Strictly speaking, because the magnetic field lines must close outside
the annulus, one needs to replace $L^{\,}_{z}$ by a length scale bounded from
below by the outer radius of the annulus.  This energy cost vanishes
as $1/R$ for $R,L^{\,}_{z}\to\infty$, which means that the ground
state degeneracy is lifted as a power law, rather than exponentially.

The reason underlying this power-law splitting is the fact that the
electromagnetic gauge field is not an emergent gauge field in the same
sense as the Chern-Simons fields that are present in, say, a
fractional topological insulator with gapped edges.  To elaborate on
this distinction, we first recall that the topological degeneracy
derived in Ref.~\onlinecite{hansson} arises from a dynamical treatment
of the electromagnetic gauge field in (2+1)-dimensional space and
time.  The topological sectors in which this degeneracy is encoded
reside in the Hilbert space of the electromagnetic gauge field, which
is in turn entangled with the Hilbert space of the electronic degrees
of freedom.  Since the photonic degrees of freedom in a real annular
superconductor also exist outside the sample, there is nothing to
prevent the environment from fixing a topological sector.  For
example, the presence of an external magnetic field in the hole can
privilege one topological sector over the other by fixing the flux
through the hole.

It is crucial to contrast this with the case of a ``true" TRS-FTL,
where the Chern-Simons fields arise naturally from electron-electron
interactions.  In this case, the topological sectors reside in the
Hilbert space of the electrons alone, and the CS fields do not exist
outside the sample.  Inserting an electromagnetic flux through the
hole of an annular TRS-FTL switches between topological sectors, but
does not betray any information about the identity of the initial or
final sector.  For this reason, the degeneracy of different
topological sectors is completely protected from the environment in
the limit of infinite system size.

\section{Summary and conclusion}

In this paper we have derived a formula for the topological ground
state degeneracy of a time-reversal symmetric, multi-component,
Abelian Chern-Simons theory. The formula, which holds when the edge
states of the theory are gapped by appropriate perturbations, says
that the GSD of the system on a planar surface with
$N^{\,}_{\mathrm{h}}$ holes is given by
$|\det\,K|^{N^{\,}_{\mathrm{h}}/2}$, where $K$ is the $K$-matrix.  We
then examined the situation where this topological degeneracy is split
exponentially by finite-size effects, and found that the set of
$N^{\,}_{\mathrm{h}}$ holes admits a description in terms of an
effective spin-like system whose couplings can be tuned by varying the
sizes and arrangement of the holes.  We also considered calorimetry as
a possible means of detecting the topological degeneracy.  The
proposed experiment would measure the contribution of the topological
degeneracy to the heat capacity at low temperatures, which we argued
could be visible on top of the expected electronic and phononic
backgrounds as long as the host material is sufficiently thin.
Finally, in light of these results, we revisited the notion that
ordinary $s$-wave superconductors are topologically ordered.  We
argued that, while thin-film superconductors do indeed possess a
ground state degeneracy on punctured planar surfaces, this degeneracy
is lifted in a power-law, rather than an exponential, fashion due to
the (3+1)-dimensional nature of the electromagnetic gauge field.

We close by pointing out several possible extensions of this work.
First, we believe that the correspondence suggested in this paper
between gluing conditions \eqref{gluing} and gapped edges of TRS-FTLs
would benefit from further study.  Sharpening this correspondence
could provide a viewpoint on fractionalized phases with gapped edges
that is complementary to the classification of such edges in terms of
Lagrangian subgroups.\
\cite{footnote-lagrangian-subgroup,Barkeshli13a,Barkeshli13b,levinPRX}
Second, we
note that our results concerning the ground state degeneracy may still
apply to TRS-FTLs where the backscattering terms of
Eq.~\eqref{tunneling} \textit{do not} respect time-reversal symmetry.
One could therefore also consider extending the results of this paper
to fractional topological insulators whose protected edge modes are
gapped by perturbations that break TRS, as is done in
Refs.~\onlinecite{lindner} and \onlinecite{motruk}.  Third, it would
be interesting to determine what other kinds of ``artificial''
spin-like systems could be realized in TRS-FTLs with more complicated
$K$-matrices than those in the class of Eq.~\eqref{class}.  It is
conceivable that remnants of the topological degeneracy may manifest
themselves as exotic properties of these less conventional models.
Finally, we must point out that a fractionalized two-dimensional state
of matter with time-reversal symmetry has not yet been discovered
experimentally, and that the search for such a state must remain a
priority.

\section*{Acknowledgments}
We are grateful to Kurt Clausen, Eduardo Fradkin, Hans Hansson,
Shivaji Sondhi, Chenjie Wang, and Frank Wilczek for enlightening discussions.
Upon completion of this work, we were made aware by Shinsei Ryu
of Ref.\ \onlinecite{wang-wen-archives12}, in which related results
were obtained.
T.I. was supported by the National Science Foundation Graduate
Research Fellowship Program under Grant No.~DGE-1247312.  T.N. was
supported by DARPA SPAWARSYSCEN Pacific N66001-11-1-4110, and C.C. was
supported by DOE Grant DEF-06ER46316.  We also acknowledge support from
the Condensed Matter Theory Visitors' Program at Boston University.

\appendix*

\section{Details on the gluing conditions \eqref{gluing}}

\subsection{Consistency conditions and constraints from TRS}

In this section, we point out various consistency conditions that
constrain the gluing conditions \eqref{gluing}. 
Let $\mathtt{i}=1,\ldots, N$.

First, let us understand why the scalar fields $\chi^{\mathtt{i}}_{\alpha}$ and
$\chi^{\mathtt{i}}_{\beta}$ are related by the same linear transformation
$T$ as are the gauge fields $\alpha^{\mathtt{i}}_{\mu}$ and
$\beta^{\mathtt{i}}_{\mu}$ appearing in Eq.~\eqref{gluing1}.  
To see this, suppose that we replace Eq.~\eqref{gluing1} by
\begin{subequations}
\begin{align}
\chi^{\mathtt{i}}_\alpha\big\vert^{\,}_{\partial D} = 
T^{\,}_\mathtt{ij}\, 
\chi^{\mathtt{j}}_{\beta}\big
\vert^{\,}_{\partial D},
\qquad
\alpha^{\mathtt{i}}_{\mu}\big\vert^{\,}_{\partial D} = 
U^{\,}_\mathtt{ij}\, 
\beta^{\mathtt{j}}_{\mu}\big\vert^{\,}_{\partial D},
\label{altgluing}
\end{align}
where $U$ and $T$ are both invertible linear transformations.  In
order for the alternative boundary conditions \eqref{altgluing} to be
well-defined, we must demand that
$U^{\,}_\mathtt{ij}\,\beta^{\mathtt{j}}_{\mu}$ transforms in the same way
under gauge transformations as $\alpha^{\mathtt{i}}_{\mu}$, i.e.,
\begin{align}
&
\alpha^{\mathtt{i}}_{\mu}\mapsto
\alpha^{\mathtt{i}}_{\mu}+\partial^{\,}_{\mu}\chi^{\mathtt{i}}_{\mu},
\\
&
U^{\,}_\mathtt{ij}\,\beta^{\mathtt{j}}_{\mu}\mapsto 
U^{\,}_\mathtt{ij}\,\beta^{\mathtt{j}}_{\mu}
+
\partial^{\,}_\mu(U^{\,}_\mathtt{ij}\,\chi^{\mathtt{j}}_{\beta})
\nonumber\\
&
\hphantom{\alpha^{\mathtt{i}}_{\mu}\mapsto}
=\alpha^{\mathtt{i}}_{\mu}
+
\partial^{\,}_\mu(U^{\,}_\mathtt{ij}\,T^{-1}_{\mathtt{jk}}\,\chi^{\mathsf{k}}_{\alpha}).
\end{align}
Equating the two expressions, we find that $U\,T^{-1}=\mathbbm{1}$, or,
equivalently, 
\begin{equation}
U=T.
\end{equation}
\end{subequations}

Next, we demonstrate that the matrix $T$ entering Eqs.~\eqref{gluing}
must have rational-valued entries in order for the bosonic edge theory
with the Lagrangian density \eqref{edge} to support point-like excitations. 
To see this, recall [c.f., e.g.,
Ref.~\onlinecite{santos}] that the bulk-edge correspondence implies that
\begin{subequations}
\begin{align}
\alpha^{\mathtt{i}}_{1}\big\vert^{\,}_{\partial D}(t,x)&=
\partial^{\,}_{x}\phi^{\,}_{\mathtt{i}}(t,x)\=:
\partial^{\,}_{x}\phi^{\mathtt{i}}_{\alpha}(t,x),
\\
\beta^{\mathtt{i}}_{1}\big\vert^{\,}_{\partial D}(t,x)&=
\partial^{\,}_{x}\phi^{\,}_{\mathtt{i}+N}(t,x)\=:
\partial^{\,}_{x}\phi^{\mathtt{i}}_{\beta}(t,x).
\end{align}
The gluing conditions \eqref{gluing1} therefore require that
\begin{align}
\partial^{\,}_{x}\,\phi^{\mathtt{i}}_{\alpha}(t,x)\=:
T^{\,}_{\mathtt{ij}}\,\partial^{\,}_{x}\phi^{\mathtt{j}}_{\beta}(t,x).
\end{align}
Integrating this equation over the whole boundary (which we take to
have length $L$) gives
\begin{align}
\phi^{\mathtt{i}}_{\alpha}(t,L)-\phi^{\mathtt{i}}_{\alpha}(t,0)=
T^{\,}_{\mathtt{ij}}\, 
[\phi^{\mathtt{j}}_{\beta}(t,L)-\phi^{\mathtt{j}}_{\beta}(t,0)].
\end{align} 
In order for the vertex operator
$\exp\big(-\mathrm{i}K^{\,}_{\mathsf{kl}}\,\bm{\phi}^{\,}_{\mathsf{l}}(t,x)\big)$
with $\mathsf{k}=1,\ldots,2N$
to obey well-defined periodic boundary conditions
(see Ref.~\onlinecite{neupert}),
\begin{align}
2\pi\,\mathbb{Z}^{2N}\ni&\,
K\[\bm{\phi}(t,L)-\bm{\phi}(t,0)\]
\nonumber\\
=&\,
\begin{pmatrix}
\kappa&\Delta\\
\Delta^{\mathsf{T}}&-\kappa
\end{pmatrix}
\begin{pmatrix}
\bm{\phi}^{\,}_{\alpha}(t,L)-\bm{\phi}^{\,}_{\alpha}(t,0)\\
\bm{\phi}^{\,}_{\beta}(t,L)-\bm{\phi}^{\,}_{\beta}(t,0)
\end{pmatrix}
\nonumber\\
=&\,
\begin{pmatrix}
\kappa&\Delta\\
\Delta^{\mathsf{T}}&-\kappa
\end{pmatrix}
\begin{pmatrix}
T\,[\bm{\phi}^{\,}_{\beta}(t,L)-\bm{\phi}^{\,}_{\beta}(t,0)]\\
\bm{\phi}^{\,}_{\beta}(t,L)-\bm{\phi}^{\,}_{\beta}(t,0)
\end{pmatrix},
\end{align}
\end{subequations}
which is only possible if the elements of the $N\times N$ matrix
$T$ are rational valued.
The vertex operators
$\exp\big(-\mathrm{i}K^{\,}_{\mathsf{kl}}\,\bm{\phi}^{\,}_{\mathsf{l}}(t,x)\big)$
then define point-like particles for $\mathsf{k}=1,\ldots,2N$.

Finally, we show that time-reversal symmetry implies the constraint
\begin{subequations}
\begin{equation}
T=
T^{-1}.  
\end{equation}
TRS (implemented by the operator $\mathcal T$) acts
on the Chern-Simons fields as
(see Ref.~\onlinecite{neupert})
\begin{align}
\alpha^{\mathtt{i}}_{\mu}(t,\bm x)\underset{\mathcal T}{\longrightarrow} 
-g^{\mu\nu}\, \beta^{\, \mathtt{i}}_{\nu}(-t,\bm x),
\end{align}
so that on the boundary Eq.~\eqref{gluing1} gives
\begin{align}
\begin{split}
\alpha^{\mathtt{i}}_{\mu}(t,\bm x)&\underset{\mathcal T}{\longrightarrow} 
-g^{\mu\nu}\, 
\beta^{\mathtt{i}}_{\nu}(-t,\bm x)
\\
&=
-g^{\mu\nu}\, 
T^{-1}_{\mathtt{ij}}\,
\alpha^{\mathtt{j}}_{\nu}(-t,\bm x).
\end{split}
\end{align}
A second application of time-reversal yields
\begin{align}
\begin{split}
\alpha^{\mathtt{i}}_{\mu}(t,\bm x)&\underset{\mathcal T^{2}}{\longrightarrow}
T^{-1}_{\mathtt{ij}}\,
\beta^{\mathtt{j}}_{\mu}(t,\bm x)
\\
&=T^{-1}_\mathtt{ij}\,T^{-1}_{\mathtt{jk}}\,\alpha^{\mathtt{k}}_{\mu}(t,\bm k).
\end{split}
\end{align}
\end{subequations}
Demanding that $\mathcal T^{2}=+1$ for the CS fields implies that
$(T^{-1})^{2}=T^{2}=\mathbbm{1}$. 

\subsection{Connection between gluing conditions and gapped edges}

In this section, we elaborate on the relationship between gluing
conditions of the form \eqref{gluing} and gapped edges of TRS-FTLs.
In particular, we show that a partial correspondence holds. Given any
matrix $T$ satisfying Eq.~\eqref{gluing2}, it is possible to
construct a gapped edge of a TRS-FTL. Conversely, given a particular
gapped edge of a TRS-FTL, it is possible to construct an appropriate
gluing condition \textit{provided} that a criterion, related to the
tunneling vectors that enter Eq.~\eqref{tunneling}, is satisfied.
While we believe that it may be possible to strengthen the latter
direction of the correspondence, we leave this for future work.

\subsubsection{Constructing a gapped edge given a gluing condition}

Suppose that we are given an invertible, $N\times N$, 
rational-valued matrix $T$ 
that satisfies Eq.~\eqref{gluing2} and respects TRS, i.e.,
it satisfies $T^{2}=\mathbbm{1}$.  We would like to construct from the
matrix $T$ a set of $N$ linearly independent vectors satisfying the
Haldane criterion \eqref{haldanecrit}.

Given such a matrix $T$, we can
construct the $2N\times N$ matrix
\begin{subequations}
\begin{equation}
\begin{pmatrix}
T\\
\pm T
\end{pmatrix}
\end{equation}
satisfying
\begin{align}\label{haldane from gluing}
\begin{pmatrix}
T^{\mathsf{T}}& \pm T^{\mathsf{T}}
\end{pmatrix}
\begin{pmatrix}
\kappa & \Delta\\
\Delta^{\mathsf{T}} & -\kappa
\end{pmatrix}
\begin{pmatrix}
T\\
\pm T
\end{pmatrix}=0.
\end{align}
\end{subequations}
Therefore, given a matrix $T$ 
(with elements $T^{\,}_{\mathtt{ij}}$, where 
$\mathtt{i},\mathtt{j}=1,\ldots,N$)
that satisfies Eq.~\eqref{gluing}, we
automatically obtain at least two sets (one for each sign of the lower
$N\times N$ block) of $N$ vectors in $\mathbb{Q}^{2N}$ that satisfy the
Haldane criterion, namely
\begin{align}
\left\{
\tilde{\bm{T}}^{\,}_{\mathtt{i}}\:=
\left(
T^{\,}_{1\mathtt{i}}\ \dots\ T^{\,}_{N\mathtt{i}}\ 
\vert \ \pm T^{\,}_{1\mathtt{i}}\ \dots\ \pm T^{\,}_{N\mathtt{i}}\ 
\right)^{\mathsf{T}}
\right\}_{\mathtt{i}=1}^{N}.
\label{Ttilde def}
\end{align}
It remains to show that we can construct from these vectors a set of
$N$ linearly independent vectors in $\mathbb Z^{2N}$ that satisfy the
Haldane criterion.  To do this, we first observe that, since the
$\tilde{T}^{\,}_{\mathtt{i}}$ are rational-valued vectors, 
we can define the rescaled set
\begin{align}
\left\{
\bm{T}^{\,}_{\mathtt{i}}\:=
m^{\,}_{\mathtt{i}}\, 
\tilde{\bm{T}}^{\,}_{\mathtt{i}}\in\mathbb{Z}^{2N}
\right\}^{N}_{\mathtt{i}=1},
\end{align}
where $m^{\,}_{\mathtt{i}}$ is the smallest integer such that 
$T^{\,}_{\mathtt{i}}\in\mathbb{Z}^{2N}$.
This rescaling can be achieved by 
\begin{equation}
T\mapsto 
T\, M,
\qquad
M\:=\text{diag}\left(m^{\,}_{1},\dots,m^{\,}_{N}\right),
\label{appeq: def rescaling with M} 
\end{equation}
which leaves Eq.~\eqref{haldane from  gluing} invariant.  
Furthermore, the rescaling does not alter the
linear dependence or independence of the set 
$\{\tilde{\bm{T}}^{\,}_{\mathtt{i}}\}^{N}_{\mathtt{i}=1}$
-- in other words, proving that the $T^{\,}_{\mathtt{i}}$ are linearly
independent for all $\mathtt{i}=1,\ldots,N$ is equivalent to proving that the 
$\tilde{\bm{T}}^{\,}_{\mathtt{i}}$ 
are linearly independent for all $\mathtt{i}=1,\ldots,N$.  To do this, we first
suppose (for contradiction) that the set 
$\{\tilde{\bm{T}}^{\,}_{\mathtt{i}}\}^{N}_{\mathtt{i}=1}$
is linearly \textit{dependent}.  This implies that there exists a set of
real numbers $\lambda^{\,}_{\mathtt{j}}$ with $\mathtt{j}=1,\dots,N$ such that
\begin{align}
\sum_{\mathtt{i}=1}^{N}
\lambda^{\,}_{\mathtt{i}}\,
\tilde{\bm{T}}^{\,}_{\mathtt{i}}=0.
\end{align}
Recalling Eq.~(\ref{Ttilde def}),
this implies in particular that
\begin{align}
\sum_{\mathtt{i}=1}^{N}
\lambda^{\,}_{\mathtt{i}}\,
\left(
T^{\,}_{1\mathtt{i}}\ \dots\ T^{\,}_{N\mathtt{i}}
\right)^{\mathsf{T}} &=0.
\end{align}
In other words, the columns of the matrix $T$ are linearly dependent.
As a result, $\det T=0$.  However, this contradicts the assumption
that $T$ is an invertible matrix.  We conclude that the set 
$\{\bm{T}^{\,}_{\mathtt{i}}\}^{N}_{\mathtt{i}=1}$
consists of $N$ linearly independent integer vectors satisfying
Haldane's criterion.

The choice of sign in the definition of the vectors 
$\tilde{\bm{T}}^{\,}_{\mathtt{i}}$
in Eq.~\eqref{Ttilde def} determines whether the tunneling processes
encoded by the vectors $\bm{T}^{\,}_{\mathtt{i}}$ conserve charge or fermion parity.
To see this, we consider contracting all of the vectors 
$\bm{T}^{\,}_{\mathtt{i}}$ 
with the charge vector $\bm{Q}$ defined in Eq.~\eqref{qdef}.  
This can be written in terms of the matrix-vector product
[recall that $M$ is defined in Eq.~(\ref{appeq: def rescaling with M})]
\begin{align}
\Big(
\begin{array}{cc}
(T\, M)^{\mathsf{T}} & \pm (T\, M)^{\mathsf{T}}
\end{array}
\Big)
\begin{pmatrix}
\varrho\\
\varrho
\end{pmatrix}
=&\,
\[(T\, M)^{\mathsf{T}}\pm (T\, M)^{\mathsf{T}}\]\varrho
\nonumber\\
=&
\begin{cases} 
2\,(T\,M)^{\mathsf{T}}\varrho,
\\
\\
0,
\end{cases}
\end{align}
if one chooses the positive or negative option, respectively.  Since the 
$N\times N$ matrix $T\, M$ has integer-valued entries, 
we conclude that the positive option conserves fermion parity 
(since $\bm{T}^{\mathsf{T}}_{\mathtt{i}}\,\bm{Q}$ is an even integer 
for any $\mathtt{i}=1,\ldots,N$), while
the negative option conserves charge 
[since $\bm{T}^{\mathsf{T}}_{\mathtt{i}}\,\bm{Q}=0$
for any $\mathtt{i}=1,\ldots,N$, 
as in Eq.~\eqref{chargeconservation}].

Furthermore, the vectors $\bm{T}^{\,}_{\mathtt{i}}$ for $\mathtt{i}=1,\ldots,N$
are by construction eigenvectors of the $2N\times 2N$ matrix
\begin{align}
\Sigma^{\,}_{1}=
\begin{pmatrix}
0&\mathbbm{1}\\
\mathbbm1 & 0
\end{pmatrix}
\end{align}
with eigenvalues $\pm 1$, so that the edge is gapped in a way that
does not explicitly break TRS.  [For an explanation of this, see the
next section, or, alternatively, Ref.~\onlinecite{neupert}.]  We
leave aside the question of whether the tunneling vectors 
$\bm{T}^{\,}_{\mathtt{i}}$
with $\mathtt{i}=1,\ldots,N$
lead to spontaneous breaking of TRS via, e.g., the mechanism pointed
out in Refs.~\onlinecite{neupert} and \onlinecite{levin_stern12}.  We
nevertheless note that the spontaneous breaking of TRS may be
unavoidable for certain choices of $K$-matrices and gluing matrices
$T$.

\subsubsection{Constructing a gluing condition given a gapped edge}

In this section we show that a gapped edge of a doubled Chern-Simons
theory implies a particular associated gluing condition, so long as an
invertibility criterion is satisfied.  

To prove this, suppose we are
given $N$ linearly-independent tunneling vectors 
$\bm{T}^{\,}_{1},\dots,\bm{T}^{\,}_{N}\in\mathbbm{Z}^{2N}$ 
that satisfy the Haldane criterion
\eqref{haldanecrit}.  Let us now build the $N\times N$ matrices
\begin{subequations}
\begin{equation} 
T\:= 
\begin{pmatrix}
(\bm{T}^{\,}_{1})^{\,}_{1}&(\bm{T}^{\,}_{2})^{\,}_{1}&\dots&(\bm{T}^{\,}_N)^{\,}_{1}\\
(\bm{T}^{\,}_{1})^{\,}_{2}&(\bm{T}^{\,}_{2})^{\,}_{2}&\dots&(\bm{T}^{\,}_N)^{\,}_{2}\\
\vdots&\vdots&\dots&\vdots\\
(\bm{T}^{\,}_{1})^{\,}_N&(\bm{T}^{\,}_{2})^{\,}_N&\dots&(\bm{T}^{\,}_N)^{\,}_N\\
\end{pmatrix}
\end{equation}
and
\begin{equation}
S^{-1}\:= 
\begin{pmatrix}
(\bm{T}^{\,}_{1})^{\,}_{N+1}&(\bm{T}^{\,}_{2})^{\,}_{N+1}&\dots&(\bm{T}^{\,}_N)^{\,}_{N+1}\\
(\bm{T}^{\,}_{1})^{\,}_{N+2}&(\bm{T}^{\,}_{2})^{\,}_{N+2}&\dots&(\bm{T}^{\,}_N)^{\,}_{N+2}\\
\vdots&\vdots&\dots&\vdots\\
(\bm{T}^{\,}_{1})^{\,}_{2N}&(\bm{T}^{\,}_{2})^{\,}_{2N}&\dots&(\bm{T}^{\,}_N)^{\,}_{2N}\\
\end{pmatrix}.
\end{equation}
\end{subequations}
As the set $\{\bm{T}^{\,}_{\mathtt{i}}\}^{N}_{\mathtt{i}=1}$ 
satisfies the Haldane criterion, then the matrices $T$ and $S^{-1}$ 
can be used to build a $2N\times N$ matrix satisfying the equation
\begin{align}
\begin{split}
0&=\begin{pmatrix}
T^{\mathsf{T}}& (S^{-1})^{\mathsf{T}}
\end{pmatrix}
\begin{pmatrix}
\kappa & \Delta\\
\Delta^{\mathsf{T}} & -\kappa
\end{pmatrix}
\begin{pmatrix}
T\\
S^{-1}
\end{pmatrix}\\
&=
T^{\mathsf{T}}\,\kappa\,T 
-
(S^{-1})^{\mathsf{T}}\,\kappa\,S^{-1}
+
T^{\mathsf{T}}\,\Delta\,S^{-1}
+
(S^{-1})^{\mathsf{T}}\,\Delta^{\mathsf{T}}\,T.
\end{split}
\label{eq: 2N times N matrix with null condition}
\end{align}
Let suppose for the moment that both $T$ and $S^{-1}$ are invertible
matrices.  If this is true, then we can multiply 
Eq.~(\ref{eq: 2N times N matrix with null condition})
on the left by $S^{\mathsf{T}}$ and on the right by $S$, to obtain
\begin{align}
(T\,S)^{\mathsf{T}}\,\kappa\, T\,S-\kappa+(T\,S)^{\mathsf{T}}\,\Delta-\Delta\,(T\,S)=0,
\end{align}
i.e.,~the matrix $TS$ exists, is invertible, and satisfies
Eq.~\eqref{gluing2}.  This invertibility requirement is the caveat
advertised at the beginning of this section.  It is unclear whether it
is possible to construct a gluing matrix with the desired properties
if this requirement is not satisfied.

Let us now impose the additional constraint that the set
$\{\bm{T}^{\,}_{\mathtt{i}}\}^{N}_{\mathtt{i}=1}$ 
of tunneling vectors does not lead to the explicit breaking of
time-reversal symmetry.  We will show that this assumption implies
that the matrix $T\,S$ satisfies the TRS condition for gluing matrices,
namely $(T\,S)^{2}=\mathbbm{1}$.  To see this, recall that time-reversal
acts on the chiral bosons $\bm{\phi}$ as
(see Ref.~\onlinecite{neupert})
\begin{subequations}
\begin{align}
\bm{\phi}(\bm x,t)\underset{\mathcal T}{\longrightarrow}
\Sigma^{\,}_{1}\, 
\bm{\phi}(\bm x,-t)
+
\pi\,K^{-1}\,\Sigma^{\,}_{\downarrow}\,\bm{Q},
\end{align}
where the $2N\times 2N$ matrices
\begin{align}
\Sigma^{\,}_{1}=
\begin{pmatrix}
0&\mathbbm{1}\\
\mathbbm{1} & 0
\end{pmatrix},
\qquad
\Sigma^{\,}_{\downarrow} = 
\begin{pmatrix}
0&0\\
0&\mathbbm{1}
\end{pmatrix}.
\end{align}
\end{subequations}
For a generic tunneling term of the form
\begin{align}
\mathcal{L}^{\,}_{\mathrm{T}}=
\sum_{\bm{T}\in\{\bm{T}^{\,}_{\mathtt{i}}\}^{N}_{\mathtt{i}=1}}
U^{\,}_{\bm{T}}(x)\, 
\cos\(\bm{T}^{\mathsf{T}}\,K\,\bm{\phi}+\zeta^{\,}_{\bm{T}}(x)\),
\end{align}
time-reversal acts as
\vskip 60 true pt
\begin{widetext}
\begin{align}
\mathcal{L}^{\,}_{\mathrm{T}}\underset{\mathcal T}{\longrightarrow}&
\sum_{\bm{T}\in\{\bm{T}^{\,}_{\mathtt{i}}\}^{N}_{\mathtt{i}=1}}
U^{\,}_{\bm{T}}(x)\, 
\cos
\Big(
\bm{T}^{\mathsf{T}}\,K\,\Sigma^{\,}_{1}\bm{\phi} 
+
\zeta^{\,}_{\bm{T}}(x)
+
\pi\,\bm{T}^{\mathsf{T}}\,\Sigma^{\,}_{\downarrow}\,\bm{Q}
\Big)
\nonumber\\
&=
\sum_{\bm{T}\in\{\bm{T}^{\,}_{\mathtt{i}}\}^{N}_{\mathtt{i}=1}}
U^{\,}_{\bm{T}}(x)\, 
\cos
\Big(
-(\Sigma^{\,}_{1}\,\bm{T})^{\mathsf{T}}\,K\,\bm{\phi} 
+
\zeta^{\,}_{\bm{T}}(x)
+
\pi\,\bm{T}^{\mathsf{T}}\,\Sigma^{\,}_{\downarrow}\,\bm{Q}
\Big)
\nonumber\\
&=
\sum_{\bm{T}\in\{\bm{T}^{\,}_{\mathtt{i}}\}^{N}_{\mathtt{i}=1}}
U^{\,}_{\bm{T}}(x)\, 
\cos
\Big(
(\Sigma^{\,}_{1}\,\bm{T})^{\mathsf{T}}\,K\,\bm{\phi} 
-
\zeta^{\,}_{\bm{T}}(x)
-
\pi\,\bm{T}^{\mathsf{T}}\,\Sigma^{\,}_{\downarrow}\,\bm{Q}
\Big)
\nonumber\\
&\overset{!}{=}
\mathcal{L}^{\,}_{\mathrm{T}}.
\end{align}
The requirement of time-reversal invariance therefore implies that,
for any $\bm{T}\in\{\bm{T}^{\,}_{\mathtt{i}}\}^{N}_{\mathtt{i}=1}$, 
there exists a $\pri{\bm{T}}\in\{\bm{T}^{\,}_{\mathtt{i}}\}^{N}_{\mathtt{i}=1}$ 
such that
\begin{align}
U^{\,}_{\pri{\bm{T}}}(x)\,
\cos
\Big(
(\pri{\bm{T}})^{\mathsf{T}}\,K\,\bm{\phi} 
+
\zeta^{\,}_{\pri{\bm{T}}}(x)
\Big)=
U^{\,}_{\bm{T}}(x)\, 
\cos
\Big(
(\Sigma^{\,}_{1}\bm{T})^{\mathsf{T}}\,
K\,
\bm{\phi} 
-
\zeta^{\,}_{\bm{T}}(x)
-
\pi\,\bm{T}^{\mathsf{T}}\,
\Sigma^{\,}_{\downarrow}\,\bm{Q}
\Big).
\end{align}
\end{widetext}
This is only possible if $\pri{\bm{T}}=\pm\bm{T}$.  [In addition, there
are constraints on the function $\zeta^{\,}_{\bm{T}}(x)$ under 
$\bm{T}\mapsto\Sigma^{\,}_{1}\,\bm{T}$ that are detailed in
Ref.~\onlinecite{neupert}.]  In other words, the set
$\{\bm{T}^{\,}_{\mathtt{i}}\}^{N}_{\mathtt{i}=1}$
of tunneling vectors must map onto itself,
possibly up to a signed permutation, under time reversal,
\begin{align}
\Sigma^{\,}_{1}
\begin{pmatrix}
T\\ S^{-1}
\end{pmatrix}=
\begin{pmatrix}
S^{-1}\\ T
\end{pmatrix}
=\begin{pmatrix}
T\,P\\ S^{-1}\,P
\end{pmatrix},
\end{align}
where $P$ is a signed permutation matrix.  (We multiply from the right
because we want to permute only the columns of $T$ and $S^{-1}$.)  The
second equality above implies that
\begin{subequations}
\begin{align}
S^{-1}&=T\,P,
\qquad
\indent T=S^{-1}\,P.
\label{T and S}
\end{align}
Observe that, since $P$ is invertible, the invertibility of $T$ is
automatic provided that $S^{-1}$ is invertible, and vice versa.
Furthermore, note that the tunneling vectors constructed in the
previous section satisfy Eq.~\eqref{T and S} (with $P=\mathbbm{1}$),
and therefore do not explicitly break TRS.  Multiplying the second
equality in Eq.~\eqref{T and S} from the right by $P$ and using the
first equality, we find that $P$ obeys
\begin{align}
S^{-1}\,P^{2}=
S^{-1},
\label{P^{2}=1}
\end{align}
\end{subequations}
which implies that $P^{2}=\mathbbm{1}$ 
if we assume that $S^{-1}$ is invertible 
(as we must in order to construct the gluing matrix $T\,S$).  
Combining this with Eq.~\eqref{T and S}, we can prove that 
$(T\,S)^{2}=\mathbbm{1}$.  
Indeed,
\begin{align}
T\,S=
S^{-1}\,P\,S\implies 
(T\,S)^{2}=
S^{-1}\,P\,S\,S^{-1}\,P\,S=
\mathbbm{1},
\end{align}
as desired.

\bibliographystyle{apsrev}

\bibliography{refs_frac_final_paper}

\begin{thebibliography}{46}
\expandafter\ifx\csname natexlab\endcsname\relax\def\natexlab#1{#1}\fi
\expandafter\ifx\csname bibnamefont\endcsname\relax
  \def\bibnamefont#1{#1}\fi
\expandafter\ifx\csname bibfnamefont\endcsname\relax
  \def\bibfnamefont#1{#1}\fi
\expandafter\ifx\csname citenamefont\endcsname\relax
  \def\citenamefont#1{#1}\fi
\expandafter\ifx\csname url\endcsname\relax
  \def\url#1{\texttt{#1}}\fi
\expandafter\ifx\csname urlprefix\endcsname\relax\def\urlprefix{URL }\fi
\providecommand{\bibinfo}[2]{#2}
\providecommand{\eprint}[2][]{\url{#2}}

\bibitem[{\citenamefont{Wen}(1989)}]{wen89}
\bibinfo{author}{\bibfnamefont{X.-G.} \bibnamefont{Wen}},
  \bibinfo{journal}{Phys. Rev. B} \textbf{\bibinfo{volume}{40}},
  \bibinfo{pages}{7387} (\bibinfo{year}{1989}).

\bibitem[{\citenamefont{Wen and Niu}(1990)}]{wen90}
\bibinfo{author}{\bibfnamefont{X.-G.} \bibnamefont{Wen}} \bibnamefont{and}
  \bibinfo{author}{\bibfnamefont{Q.}~\bibnamefont{Niu}},
  \bibinfo{journal}{Phys. Rev. B} \textbf{\bibinfo{volume}{41}},
  \bibinfo{pages}{9377} (\bibinfo{year}{1990}).

\bibitem[{\citenamefont{Wen}(1991)}]{wen91}
\bibinfo{author}{\bibfnamefont{X.-G.} \bibnamefont{Wen}},
  \bibinfo{journal}{Int. J. Mod. Phys. B} \textbf{\bibinfo{volume}{5}},
  \bibinfo{pages}{1641} (\bibinfo{year}{1991}).

\bibitem[{\citenamefont{Nayak et~al.}(2008)\citenamefont{Nayak, Simon, Stern,
  Freedman, and {Das~Sarma}}}]{nayak}
\bibinfo{author}{\bibfnamefont{C.}~\bibnamefont{Nayak}},
  \bibinfo{author}{\bibfnamefont{S.~H.} \bibnamefont{Simon}},
  \bibinfo{author}{\bibfnamefont{A.}~\bibnamefont{Stern}},
  \bibinfo{author}{\bibfnamefont{M.}~\bibnamefont{Freedman}}, \bibnamefont{and}
  \bibinfo{author}{\bibfnamefont{S.}~\bibnamefont{{Das~Sarma}}},
  \bibinfo{journal}{Rev. Mod. Phys.} \textbf{\bibinfo{volume}{80}},
  \bibinfo{pages}{1083} (\bibinfo{year}{2008}).

\bibitem[{\citenamefont{Yang and Halperin}(2009)}]{halperin}
\bibinfo{author}{\bibfnamefont{K.}~\bibnamefont{Yang}} \bibnamefont{and}
  \bibinfo{author}{\bibfnamefont{B.~I.} \bibnamefont{Halperin}},
  \bibinfo{journal}{Phys. Rev. B} \textbf{\bibinfo{volume}{79}},
  \bibinfo{pages}{115317} (\bibinfo{year}{2009}).

\bibitem[{\citenamefont{Barlas and Yang}(2012)}]{yang}
\bibinfo{author}{\bibfnamefont{Y.}~\bibnamefont{Barlas}} \bibnamefont{and}
  \bibinfo{author}{\bibfnamefont{K.}~\bibnamefont{Yang}},
  \bibinfo{journal}{Phys. Rev. B} \textbf{\bibinfo{volume}{85}},
  \bibinfo{pages}{195107} (\bibinfo{year}{2012}).

\bibitem[{\citenamefont{Cooper and Stern}(2009)}]{cooper}
\bibinfo{author}{\bibfnamefont{N.~R.} \bibnamefont{Cooper}} \bibnamefont{and}
  \bibinfo{author}{\bibfnamefont{A.}~\bibnamefont{Stern}},
  \bibinfo{journal}{Phys. Rev. Lett.} \textbf{\bibinfo{volume}{102}},
  \bibinfo{pages}{176807} (\bibinfo{year}{2009}).

\bibitem[{\citenamefont{Chickering et~al.}(2010)\citenamefont{Chickering,
  Eisenstein, Pfeiffer, and West}}]{chickering1}
\bibinfo{author}{\bibfnamefont{W.~E.} \bibnamefont{Chickering}},
  \bibinfo{author}{\bibfnamefont{J.~P.} \bibnamefont{Eisenstein}},
  \bibinfo{author}{\bibfnamefont{L.~N.} \bibnamefont{Pfeiffer}},
  \bibnamefont{and} \bibinfo{author}{\bibfnamefont{K.~W.} \bibnamefont{West}},
  \bibinfo{journal}{Phys. Rev. B} \textbf{\bibinfo{volume}{81}},
  \bibinfo{pages}{245319} (\bibinfo{year}{2010}).

\bibitem[{\citenamefont{Chickering et~al.}(2013)\citenamefont{Chickering,
  Eisenstein, Pfeiffer, and West}}]{chickering2}
\bibinfo{author}{\bibfnamefont{W.~E.} \bibnamefont{Chickering}},
  \bibinfo{author}{\bibfnamefont{J.~P.} \bibnamefont{Eisenstein}},
  \bibinfo{author}{\bibfnamefont{L.~N.} \bibnamefont{Pfeiffer}},
  \bibnamefont{and} \bibinfo{author}{\bibfnamefont{K.~W.} \bibnamefont{West}},
  \bibinfo{journal}{Phys. Rev. B} \textbf{\bibinfo{volume}{87}},
  \bibinfo{pages}{075302} (\bibinfo{year}{2013}).

\bibitem[{\citenamefont{Wen and Zee}(1992)}]{wen_zee}
\bibinfo{author}{\bibfnamefont{X.-G.} \bibnamefont{Wen}} \bibnamefont{and}
  \bibinfo{author}{\bibfnamefont{A.}~\bibnamefont{Zee}},
  \bibinfo{journal}{Phys. Rev. B} \textbf{\bibinfo{volume}{46}},
  \bibinfo{pages}{2290} (\bibinfo{year}{1992}).

\bibitem[{\citenamefont{Barkeshli et~al.}(2014)\citenamefont{Barkeshli, Oreg,
  and Qi}}]{barkeshli_oreg_qi}
\bibinfo{author}{\bibfnamefont{M.}~\bibnamefont{Barkeshli}},
  \bibinfo{author}{\bibfnamefont{Y.}~\bibnamefont{Oreg}}, \bibnamefont{and}
  \bibinfo{author}{\bibfnamefont{X.-L.} \bibnamefont{Qi}},
  \bibinfo{journal}{e-print arXiv:1401.3750}  (\bibinfo{year}{2014}).

\bibitem[{\citenamefont{Freedman et~al.}(2004)\citenamefont{Freedman, Nayak,
  Shtengel, Walker, and Wang}}]{freedman_nayak}
\bibinfo{author}{\bibfnamefont{M.}~\bibnamefont{Freedman}},
  \bibinfo{author}{\bibfnamefont{C.}~\bibnamefont{Nayak}},
  \bibinfo{author}{\bibfnamefont{K.}~\bibnamefont{Shtengel}},
  \bibinfo{author}{\bibfnamefont{K.}~\bibnamefont{Walker}}, \bibnamefont{and}
  \bibinfo{author}{\bibfnamefont{Z.}~\bibnamefont{Wang}},
  \bibinfo{journal}{Ann. Phys.} \textbf{\bibinfo{volume}{310}},
  \bibinfo{pages}{428} (\bibinfo{year}{2004}).

\bibitem[{\citenamefont{Kane and Mele}(2005{\natexlab{a}})}]{kane_mele1}
\bibinfo{author}{\bibfnamefont{C.~L.} \bibnamefont{Kane}} \bibnamefont{and}
  \bibinfo{author}{\bibfnamefont{E.~J.} \bibnamefont{Mele}},
  \bibinfo{journal}{Phys. Rev. Lett.} \textbf{\bibinfo{volume}{95}},
  \bibinfo{pages}{226801} (\bibinfo{year}{2005}{\natexlab{a}}).

\bibitem[{\citenamefont{Kane and Mele}(2005{\natexlab{b}})}]{kane_mele2}
\bibinfo{author}{\bibfnamefont{C.~L.} \bibnamefont{Kane}} \bibnamefont{and}
  \bibinfo{author}{\bibfnamefont{E.~J.} \bibnamefont{Mele}},
  \bibinfo{journal}{Phys. Rev. Lett.} \textbf{\bibinfo{volume}{95}},
  \bibinfo{pages}{146802} (\bibinfo{year}{2005}{\natexlab{b}}).

\bibitem[{\citenamefont{Bernevig and Zhang}(2006)}]{bernevig}
\bibinfo{author}{\bibfnamefont{B.~A.} \bibnamefont{Bernevig}} \bibnamefont{and}
  \bibinfo{author}{\bibfnamefont{S.-C.} \bibnamefont{Zhang}},
  \bibinfo{journal}{Phys. Rev. Lett.} \textbf{\bibinfo{volume}{96}},
  \bibinfo{pages}{106802} (\bibinfo{year}{2006}).

\bibitem[{\citenamefont{Scharfenberger
  et~al.}(2011)\citenamefont{Scharfenberger, Thomale, and Greiter}}]{thomale}
\bibinfo{author}{\bibfnamefont{B.}~\bibnamefont{Scharfenberger}},
  \bibinfo{author}{\bibfnamefont{R.}~\bibnamefont{Thomale}}, \bibnamefont{and}
  \bibinfo{author}{\bibfnamefont{M.}~\bibnamefont{Greiter}},
  \bibinfo{journal}{Phys. Rev. B} \textbf{\bibinfo{volume}{84}},
  \bibinfo{pages}{140404} (\bibinfo{year}{2011}).

\bibitem[{\citenamefont{Kitaev}(2003)}]{tc}
\bibinfo{author}{\bibfnamefont{A.~Y.} \bibnamefont{Kitaev}},
  \bibinfo{journal}{Ann. Phys.} \textbf{\bibinfo{volume}{303}},
  \bibinfo{pages}{2} (\bibinfo{year}{2003}).

\bibitem[{\citenamefont{Hansson et~al.}(2004)\citenamefont{Hansson, Oganesyan,
  and Sondhi}}]{hansson}
\bibinfo{author}{\bibfnamefont{T.~H.} \bibnamefont{Hansson}},
  \bibinfo{author}{\bibfnamefont{V.}~\bibnamefont{Oganesyan}},
  \bibnamefont{and} \bibinfo{author}{\bibfnamefont{S.~L.}
  \bibnamefont{Sondhi}}, \bibinfo{journal}{Ann. Phys.}
  \textbf{\bibinfo{volume}{313}}, \bibinfo{pages}{497} (\bibinfo{year}{2004}).

\bibitem[{\citenamefont{Levin and Stern}(2009)}]{levin_stern}
\bibinfo{author}{\bibfnamefont{M.}~\bibnamefont{Levin}} \bibnamefont{and}
  \bibinfo{author}{\bibfnamefont{A.}~\bibnamefont{Stern}},
  \bibinfo{journal}{Phys. Rev. Lett.} \textbf{\bibinfo{volume}{103}},
  \bibinfo{pages}{196803} (\bibinfo{year}{2009}).

\bibitem[{\citenamefont{Neupert et~al.}(2011)\citenamefont{Neupert, Santos,
  Ryu, Chamon, and Mudry}}]{neupert}
\bibinfo{author}{\bibfnamefont{T.}~\bibnamefont{Neupert}},
  \bibinfo{author}{\bibfnamefont{L.}~\bibnamefont{Santos}},
  \bibinfo{author}{\bibfnamefont{S.}~\bibnamefont{Ryu}},
  \bibinfo{author}{\bibfnamefont{C.}~\bibnamefont{Chamon}}, \bibnamefont{and}
  \bibinfo{author}{\bibfnamefont{C.}~\bibnamefont{Mudry}},
  \bibinfo{journal}{Phys. Rev. B} \textbf{\bibinfo{volume}{84}},
  \bibinfo{pages}{165107} (\bibinfo{year}{2011}).

\bibitem[{\citenamefont{Santos et~al.}(2011)\citenamefont{Santos, Neupert, Ryu,
  Chamon, and Mudry}}]{santos}
\bibinfo{author}{\bibfnamefont{L.}~\bibnamefont{Santos}},
  \bibinfo{author}{\bibfnamefont{T.}~\bibnamefont{Neupert}},
  \bibinfo{author}{\bibfnamefont{S.}~\bibnamefont{Ryu}},
  \bibinfo{author}{\bibfnamefont{C.}~\bibnamefont{Chamon}}, \bibnamefont{and}
  \bibinfo{author}{\bibfnamefont{C.}~\bibnamefont{Mudry}},
  \bibinfo{journal}{Phys. Rev. B} \textbf{\bibinfo{volume}{84}},
  \bibinfo{pages}{165138} (\bibinfo{year}{2011}).

\bibitem[{\citenamefont{Haldane}(1995)}]{haldane}
\bibinfo{author}{\bibfnamefont{F.~D.~M.} \bibnamefont{Haldane}},
  \bibinfo{journal}{Phys. Rev. Lett.} \textbf{\bibinfo{volume}{74}},
  \bibinfo{pages}{2090} (\bibinfo{year}{1995}).

\bibitem[{\citenamefont{Wesolowski et~al.}(1994)\citenamefont{Wesolowski,
  Hosotani, and Ho}}]{wesolowski}
\bibinfo{author}{\bibfnamefont{D.}~\bibnamefont{Wesolowski}},
  \bibinfo{author}{\bibfnamefont{Y.}~\bibnamefont{Hosotani}}, \bibnamefont{and}
  \bibinfo{author}{\bibfnamefont{C.-L.} \bibnamefont{Ho}},
  \bibinfo{journal}{Int. J. Mod. Phys. A} \textbf{\bibinfo{volume}{9}},
  \bibinfo{pages}{969} (\bibinfo{year}{1994}).

\bibitem[{\citenamefont{Wang and Levin}(2013)}]{wang}
\bibinfo{author}{\bibfnamefont{C.}~\bibnamefont{Wang}} \bibnamefont{and}
  \bibinfo{author}{\bibfnamefont{M.}~\bibnamefont{Levin}},
  \bibinfo{journal}{Phys. Rev. B} \textbf{\bibinfo{volume}{88}},
  \bibinfo{pages}{245136} (\bibinfo{year}{2013}).

\bibitem[{\citenamefont{Vestergren et~al.}(2005)\citenamefont{Vestergren,
  Lidmar, and Hansson}}]{hansson_splitting}
\bibinfo{author}{\bibfnamefont{A.}~\bibnamefont{Vestergren}},
  \bibinfo{author}{\bibfnamefont{J.}~\bibnamefont{Lidmar}}, \bibnamefont{and}
  \bibinfo{author}{\bibfnamefont{T.~H.} \bibnamefont{Hansson}},
  \bibinfo{journal}{Europhys. Lett.} \textbf{\bibinfo{volume}{69}},
  \bibinfo{pages}{256} (\bibinfo{year}{2005}).

\bibitem[{\citenamefont{Vestergren and Lidmar}(2005)}]{vestergren_prb}
\bibinfo{author}{\bibfnamefont{A.}~\bibnamefont{Vestergren}} \bibnamefont{and}
  \bibinfo{author}{\bibfnamefont{J.}~\bibnamefont{Lidmar}},
  \bibinfo{journal}{Phys. Rev. B} \textbf{\bibinfo{volume}{72}},
  \bibinfo{pages}{174515} (\bibinfo{year}{2005}).

\bibitem[{foo({\natexlab{a}})}]{footnote2}
\bibinfo{howpublished}{In the semiclassical approximation employed in Ref.\
  \onlinecite{wen90}, $\xi\sim (m^{*}\Delta)^{-1/2}$, where $m^{*}$ is the
  effective mass of the quasiparticle and $\Delta$ is the gap to quasiparticle
  excitations.}

\bibitem[{fen()}]{fendley}
\bibinfo{howpublished}{P. Fendley, J. Stat. Mech. Theor. Exp. P11020 (2012).}

\bibitem[{\citenamefont{Clarke et~al.}(2013)\citenamefont{Clarke, Alicea, and
  Shtengel}}]{clarke}
\bibinfo{author}{\bibfnamefont{D.~J.} \bibnamefont{Clarke}},
  \bibinfo{author}{\bibfnamefont{J.}~\bibnamefont{Alicea}}, \bibnamefont{and}
  \bibinfo{author}{\bibfnamefont{K.}~\bibnamefont{Shtengel}},
  \bibinfo{journal}{Nat. Commun.} \textbf{\bibinfo{volume}{4}},
  \bibinfo{pages}{1348} (\bibinfo{year}{2013}).

\bibitem[{\citenamefont{Vaezi}(2014)}]{vaeziPRX}
\bibinfo{author}{\bibfnamefont{A.}~\bibnamefont{Vaezi}},
  \bibinfo{journal}{Phys. Rev. X} \textbf{\bibinfo{volume}{4}},
  \bibinfo{pages}{031009} (\bibinfo{year}{2014}).

\bibitem[{\citenamefont{Mong et~al.}(2014)\citenamefont{Mong, Clarke, Alicea,
  Lindner, Fendley, Nayak, Oreg, Stern, Berg, Shtengel et~al.}}]{mongPRX}
\bibinfo{author}{\bibfnamefont{R.~S.~K.} \bibnamefont{Mong}},
  \bibinfo{author}{\bibfnamefont{D.~J.} \bibnamefont{Clarke}},
  \bibinfo{author}{\bibfnamefont{J.}~\bibnamefont{Alicea}},
  \bibinfo{author}{\bibfnamefont{N.~H.} \bibnamefont{Lindner}},
  \bibinfo{author}{\bibfnamefont{P.}~\bibnamefont{Fendley}},
  \bibinfo{author}{\bibfnamefont{C.}~\bibnamefont{Nayak}},
  \bibinfo{author}{\bibfnamefont{Y.}~\bibnamefont{Oreg}},
  \bibinfo{author}{\bibfnamefont{A.}~\bibnamefont{Stern}},
  \bibinfo{author}{\bibfnamefont{E.}~\bibnamefont{Berg}},
  \bibinfo{author}{\bibfnamefont{K.}~\bibnamefont{Shtengel}},
  \bibnamefont{et~al.}, \bibinfo{journal}{Phys. Rev. X}
  \textbf{\bibinfo{volume}{4}}, \bibinfo{pages}{011036} (\bibinfo{year}{2014}).

\bibitem[{\citenamefont{Garden et~al.}(2009)}]{garden_review}
\bibinfo{author}{\bibfnamefont{J.-L.} \bibnamefont{Garden}}
  \bibnamefont{et~al.}, \bibinfo{journal}{Thermochim. Acta}
  \textbf{\bibinfo{volume}{492}}, \bibinfo{pages}{16} (\bibinfo{year}{2009}).

\bibitem[{\citenamefont{Ong et~al.}(2006)\citenamefont{Ong, Bourgeois,
  Skipetrov, and Chaussy}}]{ong}
\bibinfo{author}{\bibfnamefont{F.~R.} \bibnamefont{Ong}},
  \bibinfo{author}{\bibfnamefont{O.}~\bibnamefont{Bourgeois}},
  \bibinfo{author}{\bibfnamefont{S.~E.} \bibnamefont{Skipetrov}},
  \bibnamefont{and} \bibinfo{author}{\bibfnamefont{J.}~\bibnamefont{Chaussy}},
  \bibinfo{journal}{Phys. Rev. B} \textbf{\bibinfo{volume}{74}},
  \bibinfo{pages}{140503} (\bibinfo{year}{2006}).

\bibitem[{\citenamefont{Tagliati et~al.}(2012)\citenamefont{Tagliati, Krasnov,
  and Rydh}}]{tagliati1}
\bibinfo{author}{\bibfnamefont{S.}~\bibnamefont{Tagliati}},
  \bibinfo{author}{\bibfnamefont{V.~M.} \bibnamefont{Krasnov}},
  \bibnamefont{and} \bibinfo{author}{\bibfnamefont{A.}~\bibnamefont{Rydh}},
  \bibinfo{journal}{Rev. Sci. Instrum.} \textbf{\bibinfo{volume}{83}},
  \bibinfo{pages}{055107} (\bibinfo{year}{2012}).

\bibitem[{\citenamefont{Tagliati and Rydh}(2012)}]{tagliati2}
\bibinfo{author}{\bibfnamefont{S.}~\bibnamefont{Tagliati}} \bibnamefont{and}
  \bibinfo{author}{\bibfnamefont{A.}~\bibnamefont{Rydh}}, \bibinfo{journal}{J.
  Phys.: Conf. Ser.} \textbf{\bibinfo{volume}{400}}, \bibinfo{pages}{022120}
  (\bibinfo{year}{2012}).

\bibitem[{foo({\natexlab{b}})}]{footnote3}
\bibinfo{howpublished}{If instead we used the three-dimensional Debye formula,
  we would have $C^{\rm{ph}}_{V}\sim T^{3}$, which would produce an even
  smaller contribution at low temperatures, so long as the sample is not too
  thick.}

\bibitem[{foo({\natexlab{c}})}]{footnote_disorder}
\bibinfo{howpublished}{This logarithmic correction comes from so-called
  Lifshitz tails\cite{Lifshitz} in the density of states. In two dimensions,
  localized states due to disorder provide a contribution to the density of
  states that scales as the system size $L^2$ times a function of energy that
  is exponentially suppressed in the single-particle gap. Therefore, equating
  the entropy to the logarithm of the density of states, we would expect an
  entropic contribution $\sim \log(L)$ due to these states.}

\bibitem[{foo({\natexlab{d}})}]{footnote-lagrangian-subgroup}
\bibinfo{howpublished}{A Lagrangian subgroup is defined by a set of
  mathematical properties needed to describe the condensation of
  point-particles in an Abelian Chern-Simons theory (see
  Refs.~\onlinecite{Barkeshli13a,Barkeshli13b,levinPRX}). In our context, it
  amounts to finding a maximal set of vectors $\bm{T}$ that satisfy Haldane's
  criterion and with which point-like particles can be defined through the
  construction of vertex operators. By a ``maximal set," we mean a set that
  contains $N$ $\bm{T}$ vectors when the $K$-matrix is $2N\times2N$.}

\bibitem[{\citenamefont{{Barkeshli}
  et~al.}(2013{\natexlab{a}})\citenamefont{{Barkeshli}, {Jian}, and
  {Qi}}}]{Barkeshli13a}
\bibinfo{author}{\bibfnamefont{M.}~\bibnamefont{{Barkeshli}}},
  \bibinfo{author}{\bibfnamefont{C.-M.} \bibnamefont{{Jian}}},
  \bibnamefont{and} \bibinfo{author}{\bibfnamefont{X.-L.} \bibnamefont{{Qi}}},
  \bibinfo{journal}{\prb} \textbf{\bibinfo{volume}{88}},
  \bibinfo{pages}{241103} (\bibinfo{year}{2013}{\natexlab{a}}).

\bibitem[{\citenamefont{{Barkeshli}
  et~al.}(2013{\natexlab{b}})\citenamefont{{Barkeshli}, {Jian}, and
  {Qi}}}]{Barkeshli13b}
\bibinfo{author}{\bibfnamefont{M.}~\bibnamefont{{Barkeshli}}},
  \bibinfo{author}{\bibfnamefont{C.-M.} \bibnamefont{{Jian}}},
  \bibnamefont{and} \bibinfo{author}{\bibfnamefont{X.-L.} \bibnamefont{{Qi}}},
  \bibinfo{journal}{\prb} \textbf{\bibinfo{volume}{88}},
  \bibinfo{pages}{235103} (\bibinfo{year}{2013}{\natexlab{b}}).

\bibitem[{\citenamefont{Levin}(2013)}]{levinPRX}
\bibinfo{author}{\bibfnamefont{M.}~\bibnamefont{Levin}},
  \bibinfo{journal}{Phys. Rev. X} \textbf{\bibinfo{volume}{3}},
  \bibinfo{pages}{021009} (\bibinfo{year}{2013}).

\bibitem[{\citenamefont{Lindner et~al.}(2012)\citenamefont{Lindner, Berg,
  Refael, and Stern}}]{lindner}
\bibinfo{author}{\bibfnamefont{N.~H.} \bibnamefont{Lindner}},
  \bibinfo{author}{\bibfnamefont{E.}~\bibnamefont{Berg}},
  \bibinfo{author}{\bibfnamefont{G.}~\bibnamefont{Refael}}, \bibnamefont{and}
  \bibinfo{author}{\bibfnamefont{A.}~\bibnamefont{Stern}},
  \bibinfo{journal}{Phys. Rev. X} \textbf{\bibinfo{volume}{2}},
  \bibinfo{pages}{041002} (\bibinfo{year}{2012}).

\bibitem[{\citenamefont{Motruk et~al.}(2013)\citenamefont{Motruk, Berg, Turner,
  and Pollmann}}]{motruk}
\bibinfo{author}{\bibfnamefont{J.}~\bibnamefont{Motruk}},
  \bibinfo{author}{\bibfnamefont{E.}~\bibnamefont{Berg}},
  \bibinfo{author}{\bibfnamefont{A.~M.} \bibnamefont{Turner}},
  \bibnamefont{and} \bibinfo{author}{\bibfnamefont{F.}~\bibnamefont{Pollmann}},
  \bibinfo{journal}{Phys. Rev. B} \textbf{\bibinfo{volume}{88}},
  \bibinfo{pages}{085115} (\bibinfo{year}{2013}).

\bibitem[{\citenamefont{{Wang} and {Wen}}(2012)}]{wang-wen-archives12}
\bibinfo{author}{\bibfnamefont{J.}~\bibnamefont{{Wang}}} \bibnamefont{and}
  \bibinfo{author}{\bibfnamefont{X.-G.} \bibnamefont{{Wen}}},
  \bibinfo{journal}{e-print arXiv:1212.4863}  (\bibinfo{year}{2012}).

\bibitem[{\citenamefont{Levin and Stern}(2012)}]{levin_stern12}
\bibinfo{author}{\bibfnamefont{M.}~\bibnamefont{Levin}} \bibnamefont{and}
  \bibinfo{author}{\bibfnamefont{A.}~\bibnamefont{Stern}},
  \bibinfo{journal}{Phys. Rev. B} \textbf{\bibinfo{volume}{86}},
  \bibinfo{pages}{115131} (\bibinfo{year}{2012}).

\bibitem[{\citenamefont{Lifshitz}(1964)}]{Lifshitz}
\bibinfo{author}{\bibfnamefont{I.~M.} \bibnamefont{Lifshitz}},
  \bibinfo{journal}{Adv.~Phys.} \textbf{\bibinfo{volume}{13}},
  \bibinfo{pages}{483} (\bibinfo{year}{1964}).

\end{thebibliography}

\end{document}